\documentclass[conference]{IEEEtran}

\usepackage{algorithm}
\usepackage{algorithmicx}
\usepackage[noend]{algpseudocode}
\algnewcommand\algorithmicforeach{\textbf{for each}}
\algdef{S}[FOR]{ForEach}[1]{\algorithmicforeach\ #1\ \algorithmicdo}

\usepackage{booktabs} 
\usepackage[english]{babel}
\usepackage[utf8]{inputenc}
\usepackage{algorithm} 
\usepackage[noend]{algpseudocode}
\usepackage{graphicx}
\usepackage{graphics}
\usepackage{xspace}
\usepackage{subfigure}
\usepackage{float}
\usepackage{xcolor, soul}
\usepackage{setspace}
\usepackage{makecell}
\usepackage{url}
\usepackage{breakurl}

\usepackage{makecell}
\usepackage{multirow}
\usepackage{multicol}
\usepackage{array, caption, tabularx,  ragged2e,  booktabs}
\usepackage[a-1b]{pdfx}
\usepackage{amsmath}
\usepackage{listings}
\usepackage{enumitem}

\AtBeginDocument{%
  \providecommand\BibTeX{{%
    \normalfont B\kern-0.5em{\scshape i\kern-0.25em b}\kern-0.8em\TeX}}}

\begin{document}
\title{I/O Burst Prediction for HPC Clusters using Darshan Logs}

\author{\IEEEauthorblockN{Ehsan Saeedizade}
\IEEEauthorblockA{University of Nevada, Reno\\
ehsansaeedizade@nevada.unr.edu}
\and \IEEEauthorblockN{Roya Taheri}
\IEEEauthorblockA{University of Nevada, Reno\\
rtaheri@nevada.unr.edu}
 \and \IEEEauthorblockN{Engin Arslan}
    \IEEEauthorblockA{University of Texas at Arlington\\ engin.arslan@uta.edu}
}

\maketitle              
\begin{abstract}
Understanding cluster-wide I/O patterns of large-scale HPC clusters is essential to minimize the occurrence and impact of I/O interference. Yet, most previous work in this area focused on monitoring and predicting task and node-level I/O burst events. This paper analyzes Darshan reports from three supercomputers to extract system-level read and write I/O rates in five minutes intervals. We observe significant (over $100\times$) fluctuations in read and write I/O rates in all three clusters. We then train machine learning models to estimate the occurrence of system-level I/O bursts $5-120$ minutes ahead. Evaluation results show that we can predict I/O bursts with more than $90$\% accuracy (F-1 score) five minutes ahead and more than $87$\% accuracy two hours ahead. We also show that the ML models attain more than $70\%$ accuracy when estimating the degree of the I/O burst. We believe that high-accuracy predictions of I/O bursts can be used in multiple ways, such as postponing delay-tolerant I/O operations (e.g., checkpointing), pausing nonessential applications (e.g., file system scrubbers), and devising I/O-aware job scheduling methods. To validate this claim, we simulated a burst-aware job scheduler that can postpone the start time of applications to avoid I/O bursts. We show that the burst-aware job scheduling can lead to an up to  $5\times$ decrease in application runtime.



\end{abstract}

\section{Introduction}
As the scale of data generated by scientific applications increases rapidly, I/O is becoming increasingly important to store and process the gathered data on time. HPC clusters utilize large-scale high-throughput parallel file systems such as Lustre and BeeGFS to meet this demand. Yet, the shared nature of file systems often results in I/O interference, thereby significantly increasing the execution time of applications~\cite{patel2019revisiting,kim2020towards}. To demonstrate this, we measured the execution times of Block Tri-diagonal (BT)~\cite{NASParallelBenchmark} and Montage~\cite{jacob2009montage} applications with and without I/O interference using on a server with $4$ Intel Core i5 CPUs, $16$ GiB memory, and one $256$ GiB SATA SSD disk.

\begin{figure*}[t]
\begin{center}
    \subfigure[BT Benchmark]{
    \includegraphics[keepaspectratio=true,angle=0,width=.32\textwidth]{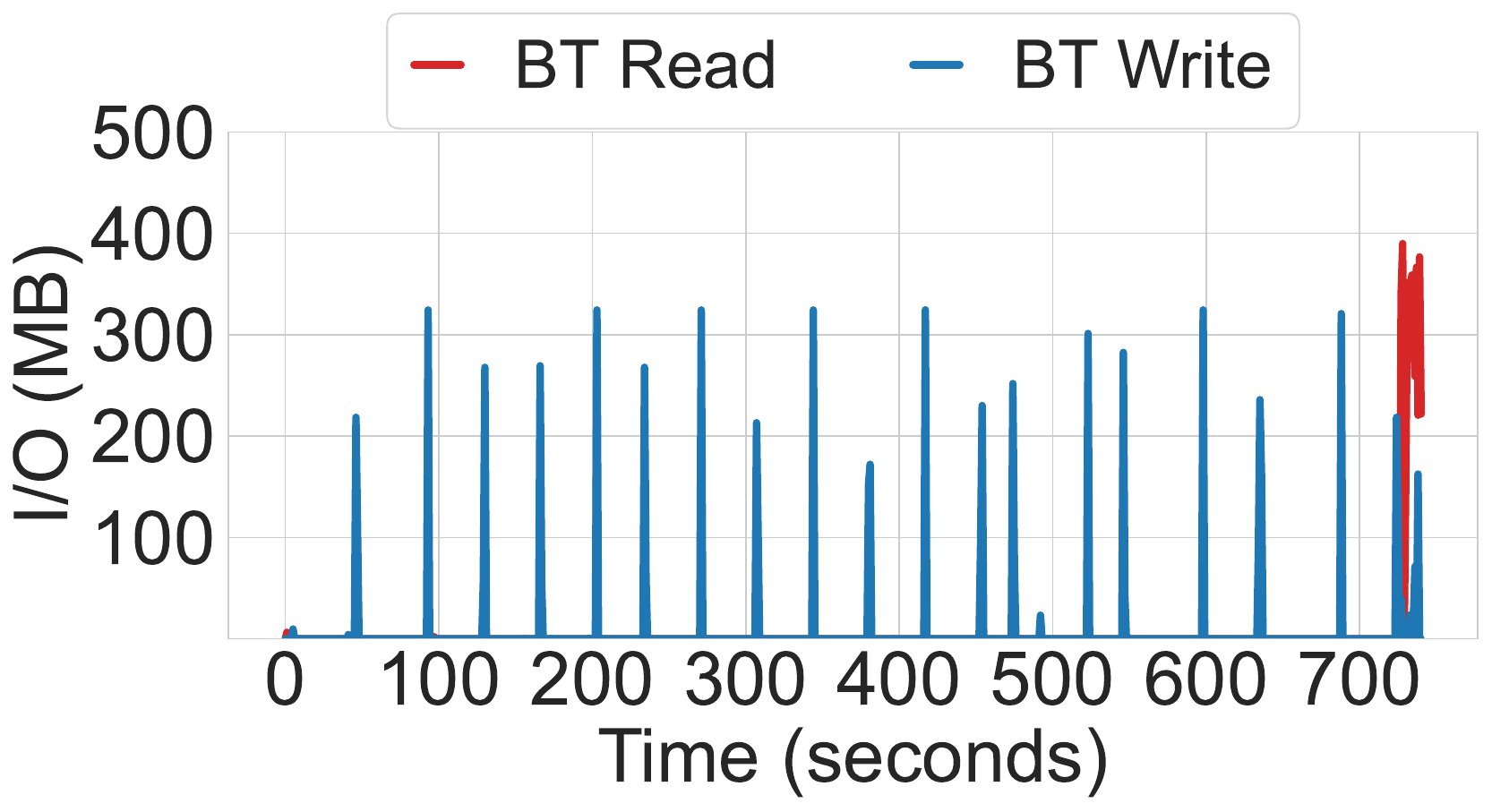}
    \label{fig:BT_IO}}
    \hspace{-4mm}
    \subfigure[Montage]{
    \includegraphics[keepaspectratio=true,angle=0,width=.32\textwidth]{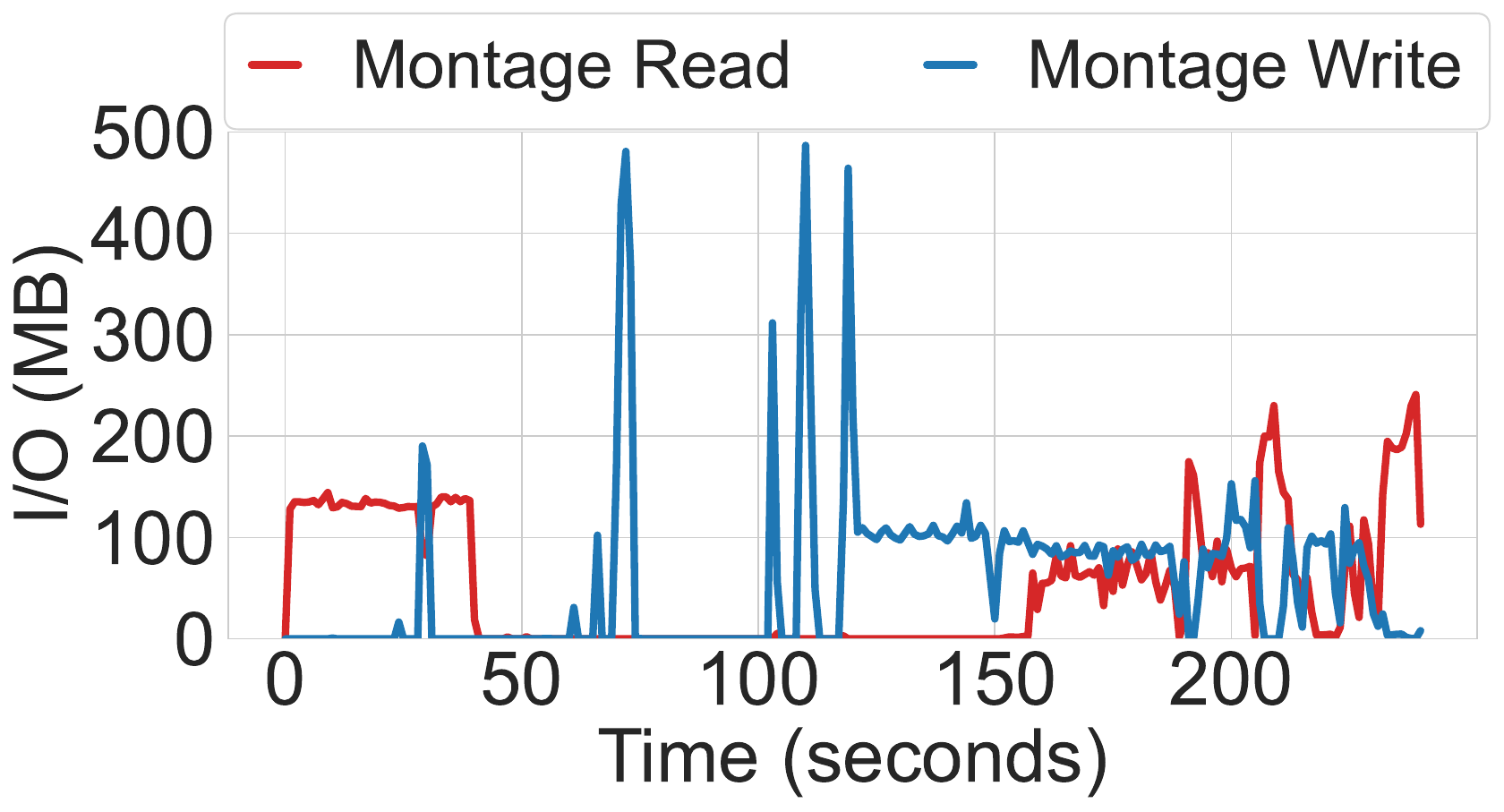}
    \label{fig:Montage_IO}}
    \hspace{-4mm}
    \subfigure[Execution times with I/O burst]{
    \includegraphics[width=.28\textwidth]{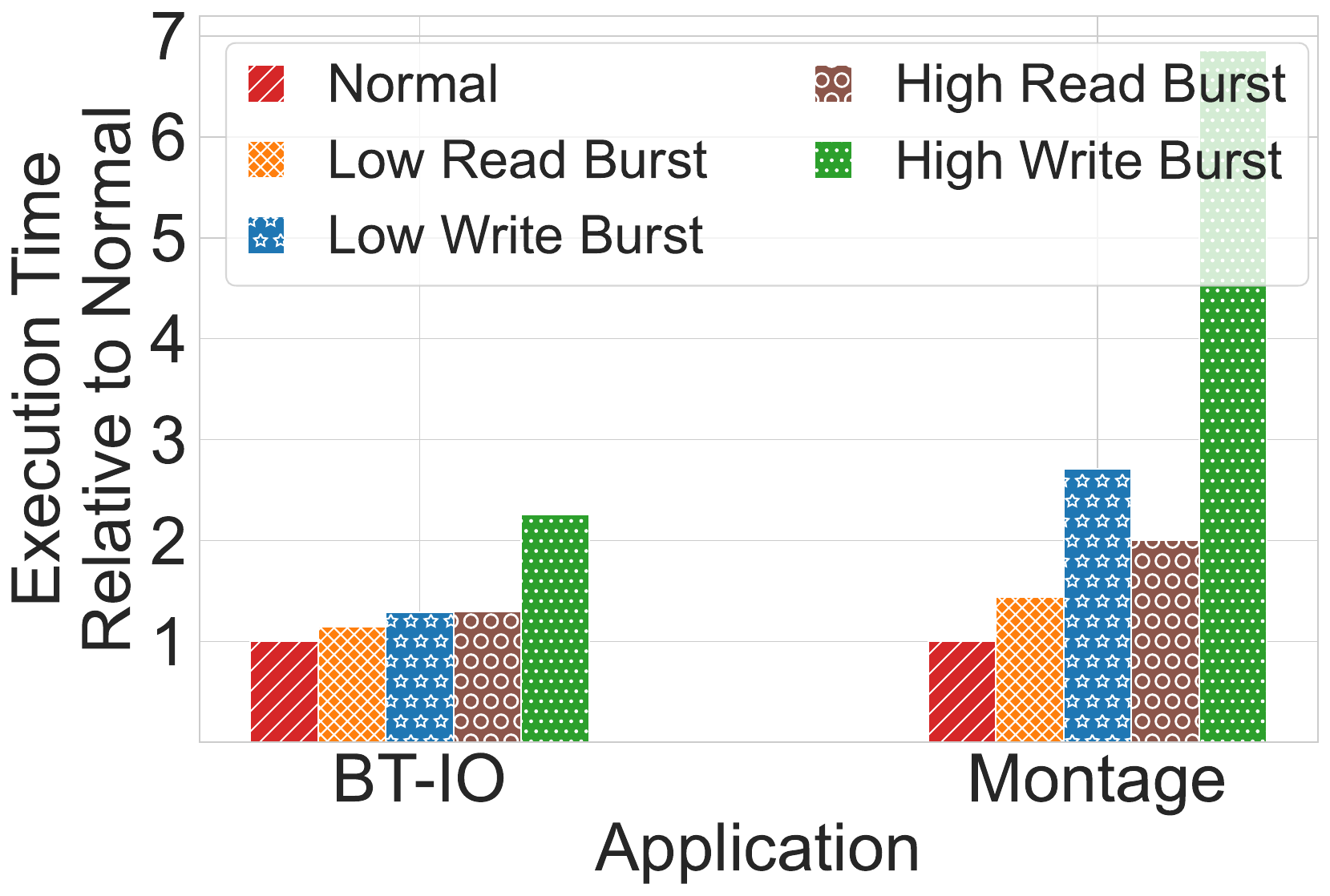} 
    \label{fig:app_execution_time}
    }
    \vspace{-1mm}
   
\caption{I/O access patterns and execution times under different background I/O load for BT-IO Benchmark and Montage applications.}
\vspace{-4mm}
\label{fig:AppS_IO_per_time}
\end{center}
\end{figure*}

BT is a NAS Parallel Benchmark application representing HPC jobs that utilize collective buffering for I/O. Montage application, however, is used to stitch together massive sky mosaics. It is characterized as a data-intensive application due to issuing a large volume of I/O operations. Figure~\ref{fig:Montage_IO} shows read and write operations issued during its execution when it is used to aggregate five data cubes, each in size of $1.07$ GiB, into a mosaic. To simulate I/O bursts, we run one (low burst) or three (high burst) background threads that continuously read or write files from/to the file system while the BT and Montage jobs are running. Figure~\ref{fig:app_execution_time} demonstrates the impact of read and write I/O burst on the execution time of the applications relative to a no-burst scenario. BT's execution time increases by $1.14\times$ and $1.28\times$ for low read and write I/O congestion and  $1.29\times$, $2.25\times$ for high read and write I/O congestion. Montage, on the other hand, is affected more due to issuing a higher level of I/O as its execution time is increased by $1.44\times$ and $2.01\times$ as a result of read I/O congestion and $2.7\times$, $6.8\times$ due to write I/O congestion. Consequently, avoiding I/O bursts is crucial to minimize the performance fluctuations of HPC applications.


Previous studies in this area mainly focused on predicting application-level I/O characteristics~\cite{wang2018iominer,gainaru2015scheduling} to tune file system configurations such as stripe size, strip count, and prefetching strategy~\cite{kim2019dca,yang2022end}. 
This work fills the gap by providing system-level I/O burst prediction only using Darshan logs. To do so, we processed nearly five million Darshan log files from three supercomputers, Blue Waters~\cite{bluewaters}, Mira~\cite{Mira}, and Theta~\cite{Theta_2}. We then analyze the occurrence of read-and-write I/O bursts and trained machine learning models. After evaluating various feature sets and machine learning models, we find that the XGBoost achieves $89-97\%$ accuracy (based on F-score) when predicting I/O bursts $5-120$ minutes ahead. Since the severity of the burst affects the degree of the impact on application performance, we also trained models to predict the burst level (based on five burst classes). XGBoost model, again, attained the best performance with $70-86\%$ accuracy.An ability to accurately predict the occurrence of I/O bursts in advance can be used for various purposes, such as scheduling large-scale I/O operations to minimize interference. For example, if an I/O burst is expected to worsen within the next $5$ minutes, delay-tolerant I/O operations (e.g., checkpointing) can be scheduled before or after the expected I/O burst to minimize the impact of I/O bursts. 

The contributions of the paper can be summarized as follows:

\begin{enumerate}
    \item We processed nearly five million Darshan log files collected at Blue Waters, Mira, and Theta supercomputers between 2014 and 2022 to extract system-level read and write I/O behavior in $5-120$ minute intervals. We show that I/O bursts do not follow a predictable pattern, making it difficult to predict using simple methods.
    \item We identified several features from Darshan logs and trained multiple machine learning models to forecast I/O bursts. We find that XGBoost models achieve the best overall performance when estimating the occurrence of I/O bursts. They attain more than $89\%$ accuracy (based on F-1 score) when predicting I/O bursts five minutes in advance. Moreover, it can achieve more than $70\%$ when predicting the severity level of I/O bursts.
    \item We developed a job scheduler simulation to demonstrate the benefits of delaying job execution times to minimize exposure to I/O bursts. We measure that application runtimes can be reduced by up to $5\times$ by delaying the start time of the application to minimize the overlap with bursty time intervals.
\end{enumerate}


\definecolor{super_light_gray}{rgb}{0.90,0.97, 0.97}
\begin{lstlisting}[belowskip=-1\baselineskip,frame=single,xleftmargin=.03\textwidth, xrightmargin=.03\textwidth,tabsize=1, breaklines=true, caption=Sample Darshan log, label={lst:darshan_header}, float=*, basicstyle=\footnotesize, backgroundcolor=\color{super_light_gray}]
# uid: 336263			
# jobid: 7738588			
# start_time: 1509271803			
# start_time_asci: Sun Oct 29 03:10:03 2017			
# end_time: 1509355904			
# end_time_asci: Mon Oct 30 02:31:44 2017			
# nprocs: 48			
# run time: 84102
#<module>	<rank>	<record id>	<counter>	<value>	<file name>	<mount pt>	<fs type>"
POSIX	0	129625958266154176	POSIX_READS	8409	/mnt/c/1218605708	/mnt/c	lustre
POSIX	0	129625958266154176	POSIX_BYTES_READ	5924503	/mnt/c/1218605708	/mnt/c	lustre
POSIX	0	129625958266154176	POSIX_F_READ_START_TIMESTAMP	15924.384187	/mnt/c/1218605708	/mnt/c	lustre
POSIX	0	129625958266154176	POSIX_F_READ_END_TIMESTAMP	77689.803174	/mnt/c/1218605708	/mnt/c	lustre
\end{lstlisting}


\section{Background}
\textbf{HPC Clusters:} \textit{Blue Waters} operated between 2012-2021 and had 22,636 CPU and 4,228 GPU nodes~\cite{bluewaters}. It was equipped with a Lustre parallel file system with a total capacity of $25$ PiB. Darshan logs for more than two million jobs are collected between 2015 and 2019~\cite{bluewaters}. \textit{Mira}, located at the Argonne Leadership Computing Facility (ALCF), was an IBM Blue Gene/Q system with $49,152$ CPU nodes and $25$ PiB storage capacity based on GPFS~\cite{Mira}. It operated between 2012 and 2019~\cite{luu2015multiplatform}. Nearly $700K$ Darshan files were collected from Mira from January 2014 to December 2020~\cite{ALCF_Public_Data}. Finally, \textit{Theta}, also hosted at ALCF is a Cray system, with $4,392$ CPU and $24$ GPU \cite{Theta_2}. Theta uses Lustre as a file system and is equipped with $10$ PiB of storage capacity. Darshan logs from more than two million jobs are collected between July 2017 to December 2022.

\textbf{Data Collection:} \textit{Darshan} is a profiling tool used to track the I/O operations of HPC applications on a sub-process level. The logs contain information on Portable Operating System Interface (POSIX) counters and Lustre file operations, among others. Files are recorded by job and time period, but the length of the time period is dependent on the frequency of operations for the job. Listing~\ref{lst:darshan_header} shows a sample Darshan file that contains metadata information for a job as well as the format of I/O records. We identify the amount and size of read/write operations by parsing \texttt{POSIX\_READS} and \texttt{POSIX\_BYTES\_READ}. Then, we use \texttt{START\_TIMESTAMP} and \texttt{END\_TIMESTAMP} fields to determine the time interval of the I/O operations. If an I/O operation spans multiple time intervals, then we split the I/O size to cover intervals equally. By processing all occurrences of a variety of counters within a file, the overall read and write I/O activity for a job can be calculated. 
Darshan logs from the Blue Waters cluster include I/O statistics in file-level precision for each process whereas they only include process-level I/O statistics for Mira and Theta clusters. File-level precision helps us locate the time interval that I/O operations issued more precisely compared to process-level. 

\section{Related work}
Wan et al. developed an I/O monitoring framework that periodically collects I/O performance metrics and job status trace on two supercomputer systems~\cite{wan2020performance}. The gathered traces are then used to predict I/O bursts for each storage server (i.e., Lustre OSS) using various machine learning models, including Regression Trees, Naive Bayes, Gradient Boosting, Support Vector Machines, Random Forest, and Neural Networks. The models can predict write bursts with a $70-90$\% accuracy. While storage node-level burst prediction is helpful, it requires additional data collection from the storage system, compute nodes, and job scheduler, limiting the applicability. On the other hand, our solution relies on a widely available Darshan tool; thus, it can be widely adopted without installing a new monitoring framework. Moreover, we present prediction models that can forecast the occurrence of an I/O burst up to two hours earlier.

Kim et al. analyzed the relationship between system logs and applications' I/O behavior and the overall I/O performance of the Cori cluster~\cite{kim2020towards}. They then identified the features that impact the I/O performance from Slurm and Darshan and Lustre Monitoring Tool \cite{LMT} logs and used regression models to predict the I/O performance with up to $84\%$ prediction accuracy.
In \cite{kunkel2015predicting}, a decision tree-based model is applied to capture and predict climate applications’ non-contiguous I/O behaviors using the SIOX framework.
In another work, supervised machine learning models are trained using long-term job traces collected by the job scheduler and Lustre monitoring tools. The gathered data is used to predict application runtime and I/O traffic pattern of batch jobs~\cite{mckenna2016machine}. They show that the Neural Network model can accurately predict the runtime of $73\%$ of jobs. The authors in \cite{schmidt2016predicting} trained Neural Network models to predict the file access times. The authors collected training data using a synthetic benchmark that issues POSIX-based random and sequential read/write operations. While all of these studies focus on predicting the I/O behavior of individual applications, this work presents a solution to estimate the system-level I/O utilization of HPC clusters.

\begin{figure}
\centering
\includegraphics[width=1\linewidth]{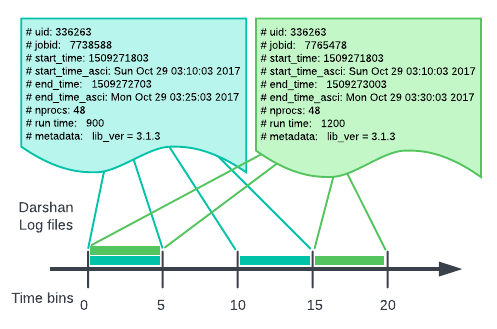} 
    \caption{Visual representation of Darshan log processing to measure system-level I/O rates in 5-minute intervals. Read and Write I/O are processed separately.}
    \label{fig:darshan_log_to_bins}
    \vspace{-5mm}
\end{figure}

Several other studies characterized the I/O performance variability in HPC systems to develop I/O interference mitigation strategies.
Herbein et al. modeled available link bandwidth between different storage layers (i.e., burst buffers, I/O network, and PFS) and developed an I/O-aware scheduler~\cite{herbein2016scalable}. The proposed scheduler reduces job performance variability by up to $33\%$ and decreases I/O-related utilization losses by up to $21\%$. Yang et al. proposed AIOT, an auto-tuning method for multi-layer storage systems that predicts the job’s I/O patterns and utilization of the storage resources for the Sunway TaihuLight supercomputer. It adopts a best-effort job scheduling scheme based on the I/O pattern and the real-time system workload~\cite{yang2022end}. While AIOT predicts I/O patterns of applications based on historical data, this study estimates system-level I/O patterns using Darshan logs.
Another study investigates the temporal, spatial, and correlative behavior of  the NERSC HPC I/O by analyzing different storage system components (e.g., Lustre, OSTs, OSSes, and MDSes). It uses a statistical characterization methodology to show that write I/O is more bursty than read in HPC I/O \cite{patel2019revisiting}.
\cite{gainaru2015scheduling} introduced a global I/O scheduler that minimizes I/O congestion by considering applications' past I/O behavior and system characteristics when scheduling I/O requests. This scheduler reduces I/O delays incurred by each application and increases overall system throughput by up to $56\%$.
\cite{lux2018predictive}~proposed a model to predict the maximum I/O performance of HPC clusters using a four-dimensional dataset collected by the IOzone benchmark.
\section{Read/Write I/O Analysis}
\label{section:Experiment_Setup}

In this section, we analyze system-wide I/O behavior and define I/O bursts. Figure \ref{fig:darshan_log_to_bins} shows how the I/O operations from Darshan logs are packed into time bins. We grouped them into 5-minute intervals so that all I/O executed within the same interval are aggregated to find the cluster-level I/O usage. Since applications can run for hours or days, we process each Darshan log line-by-line to identify when and how much an application issued I/O operations. Then, we identify the time bins that an I/O operation covers and increment the total I/O size of each bin by the size of the I/O operation divided by the number of time bins. For the sake of simplicity, we assumed that I/O activity was spread evenly over the time period of each sub-process within each file. By iterating through the entire time period, we were able to estimate the overall system I/O rates in five minutes precision. This resulted in nearly $690K$, $640K$, and $410K$ data points for Blue Waters, Mira, and Theta clusters, respectively. 


\begin{figure*} [t]
\begin{center}
    \hspace{-3mm}
    \subfigure[Hourly Read Rates]{ \includegraphics[keepaspectratio=true,angle=0,width=.31\textwidth]{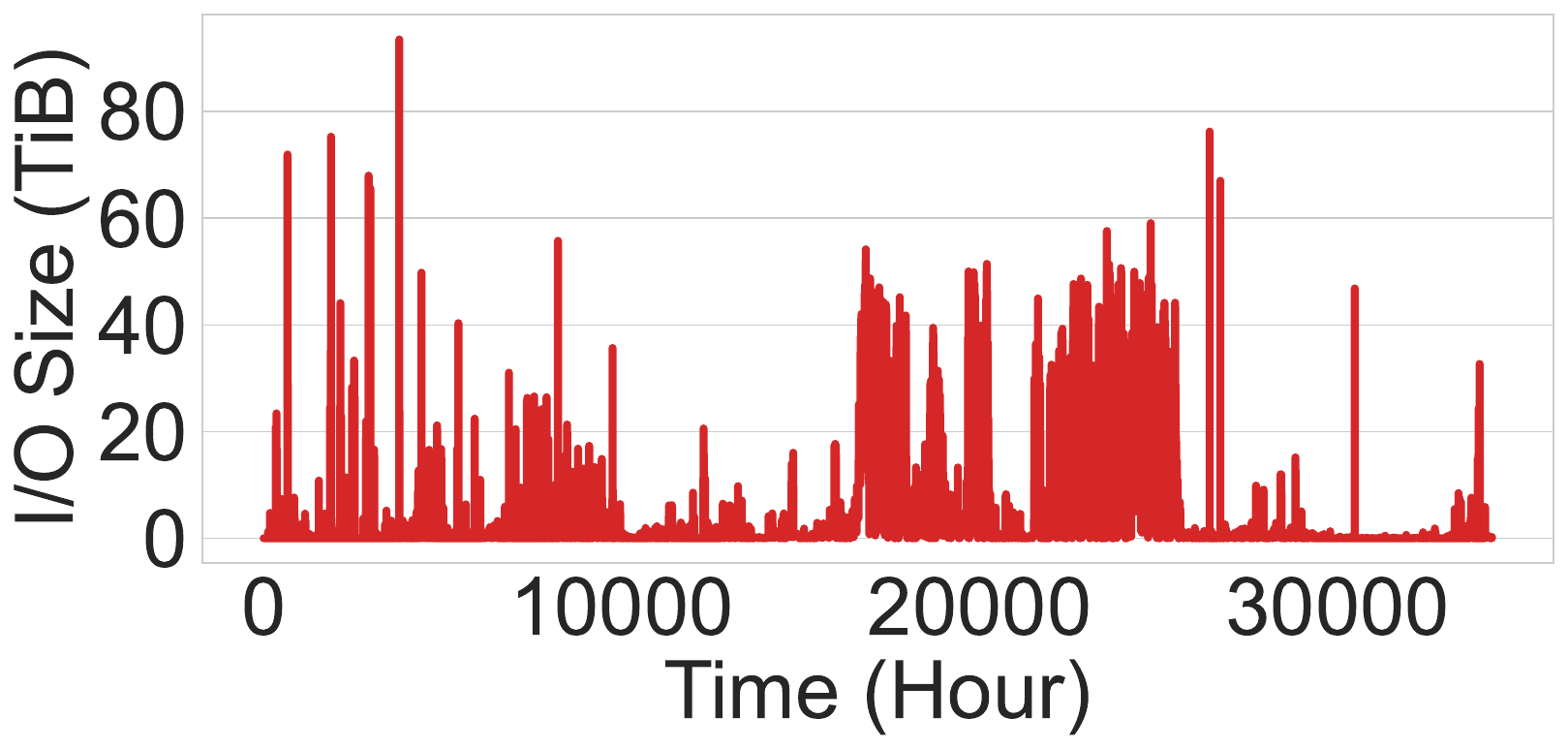}
    \label{fig:read_per_hour}}
    \hspace{-3mm}
    \subfigure[Daily Read Rates]{ \includegraphics[keepaspectratio=true,angle=0,width=.31\textwidth]{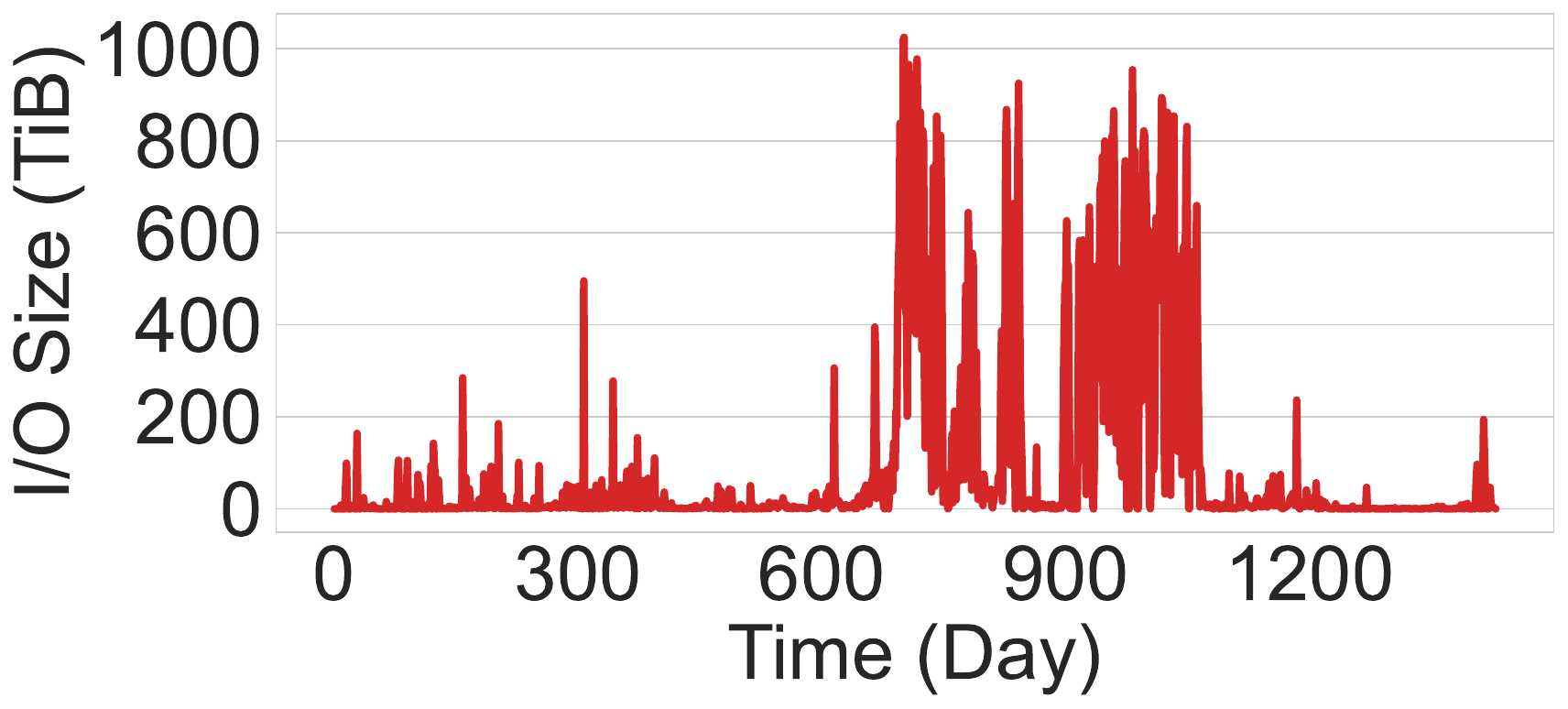}
    \label{fig:read_per_day}}
    \hspace{-3mm}
    \subfigure[Monthly Read Rates]{
\includegraphics[keepaspectratio=true,angle=0,width=.31\textwidth]{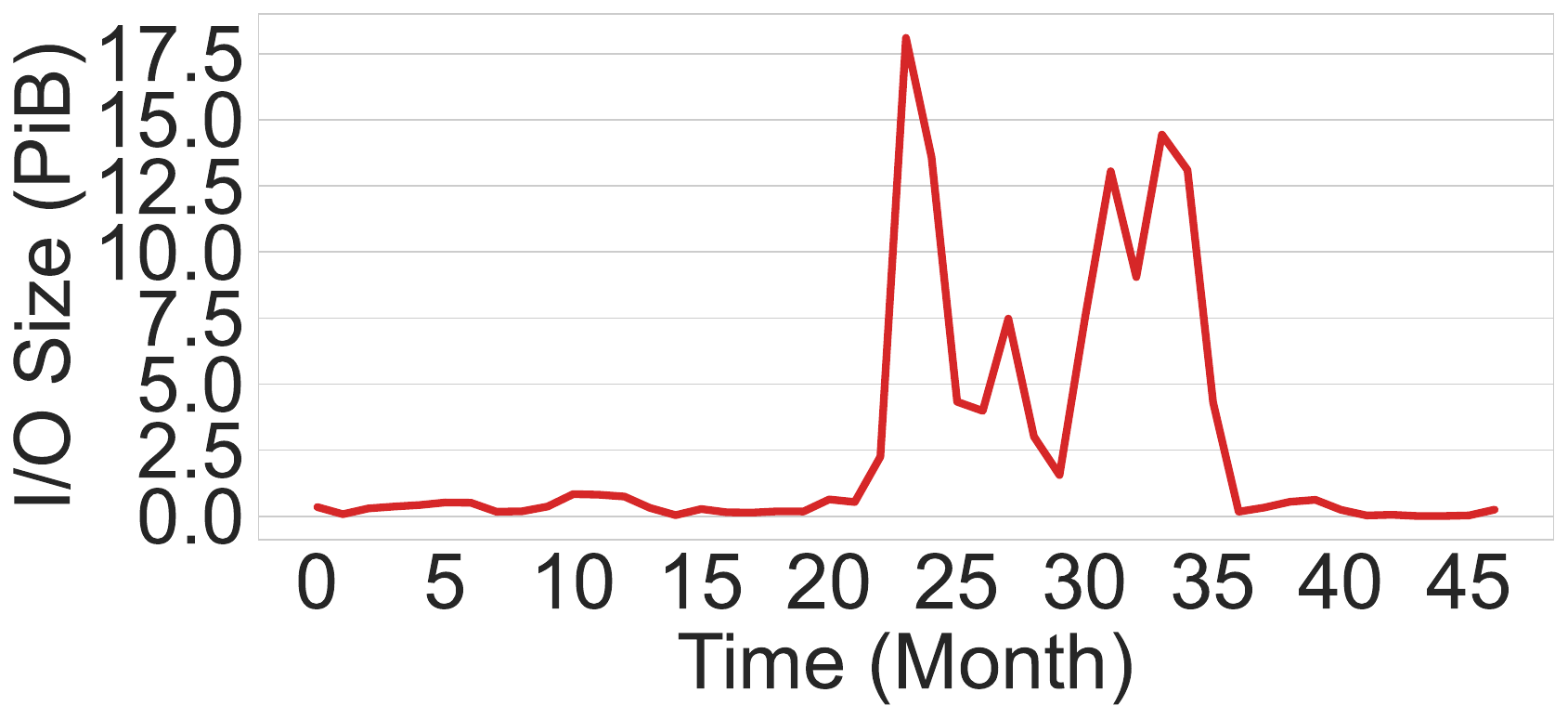}
    \label{fig:read_per_month}}
    \hspace{-3mm}
    \subfigure[Hourly Write Rates]{ \includegraphics[keepaspectratio=true,angle=0,width=.31\textwidth]{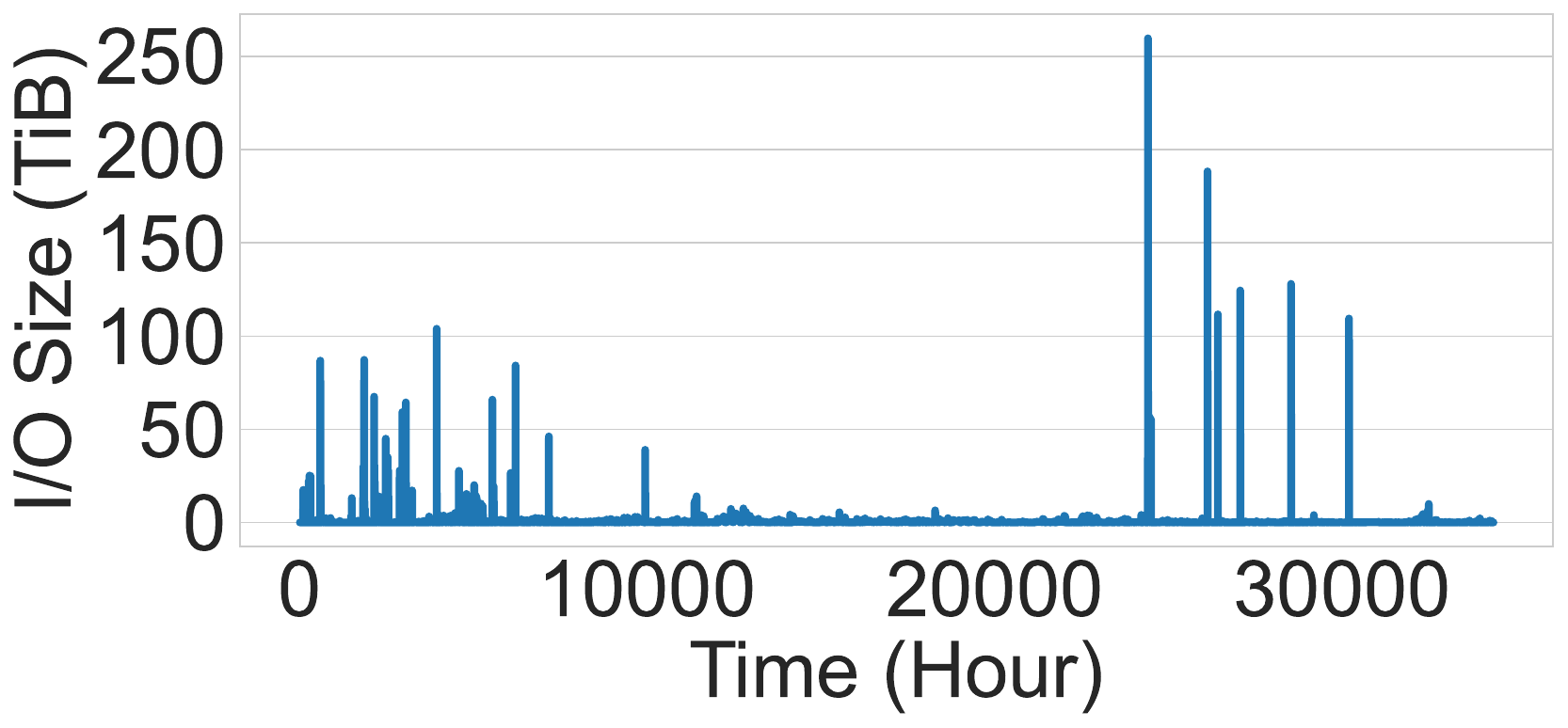}
    \label{fig:write_per_hour}}
    \hspace{-3mm}
    \subfigure[Daily Write Rates]{ \includegraphics[keepaspectratio=true,angle=0,width=.31\textwidth]{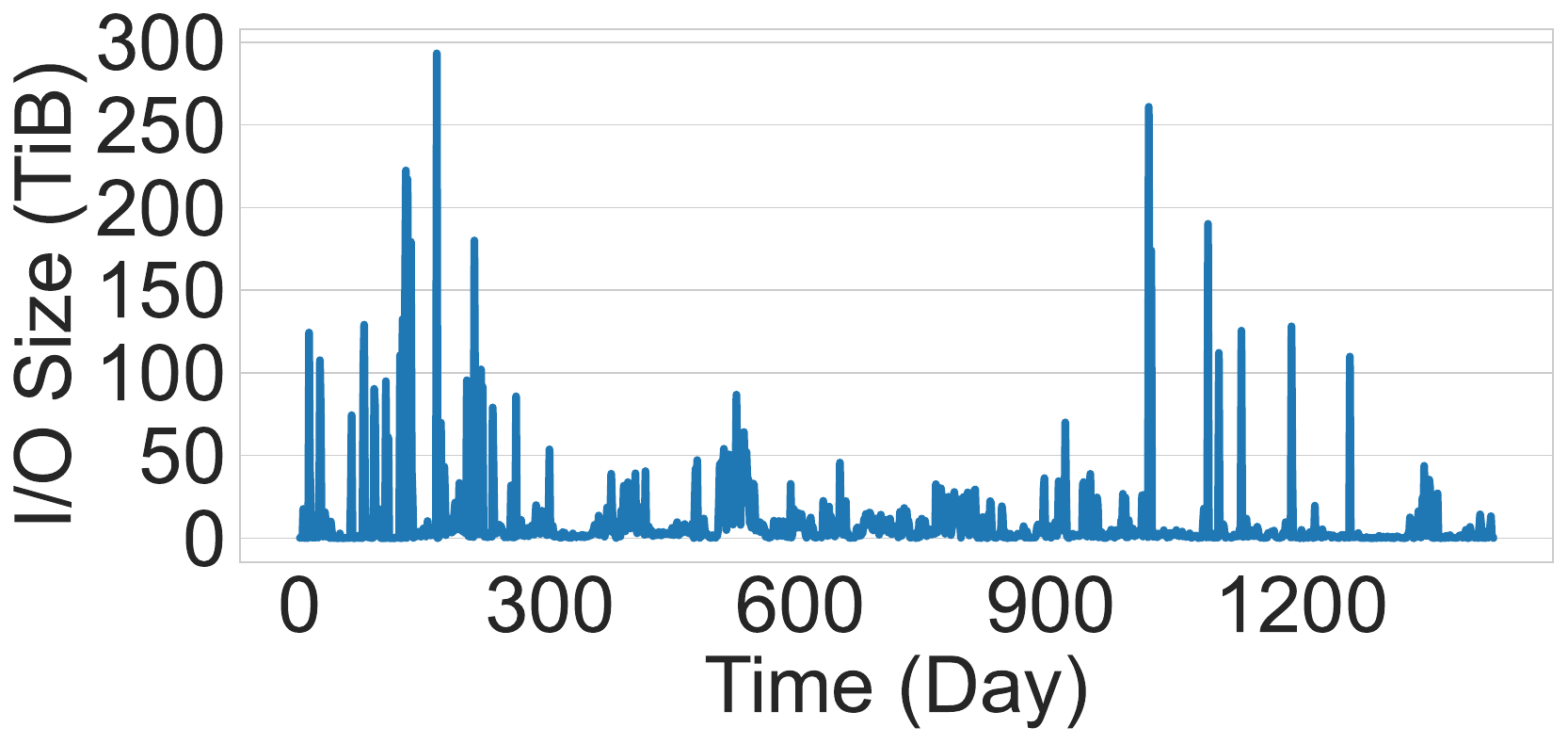}
    \label{fig:write_per_day}}
    \hspace{-3mm}
    \subfigure[Monthly Write Rates]{
\includegraphics[keepaspectratio=true,angle=0,width=.31\textwidth]{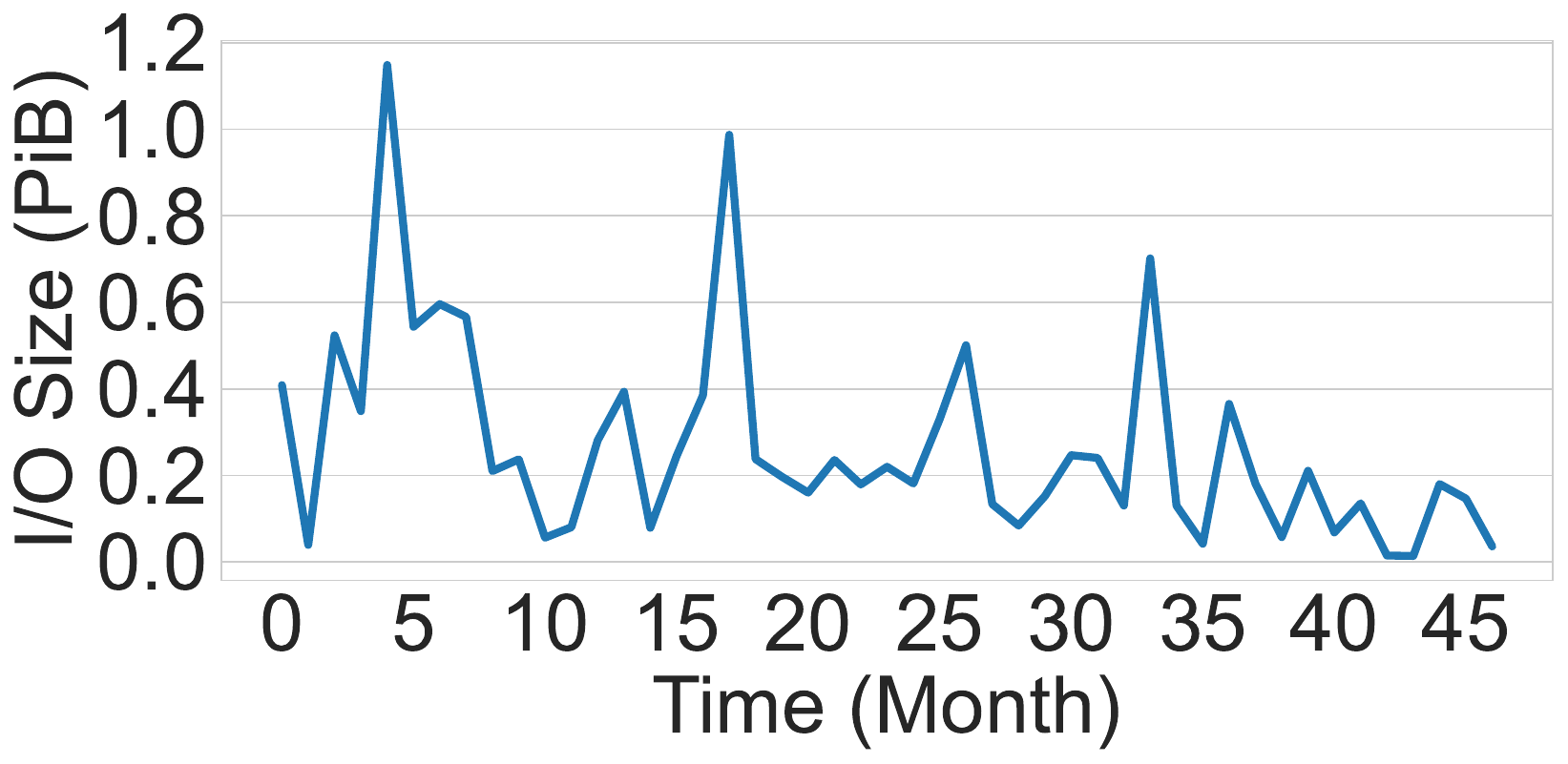}
    \label{fig:write_per_month}}
\vspace{-3mm} 
\caption{Analysis of system-level read/write I/O rates. We observe significant ($10-100x$) fluctuations in I/O rates both in hourly and monthly rates.}
\vspace{-8mm}
\label{fig:IO_per_time}
\end{center}
\end{figure*}

Figure~\ref{fig:IO_per_time} presents the total read and write operations on an hourly, daily, and monthly basis for Blue Waters. We observe that read I/O is typically larger than write I/O. Moreover, read I/O rates increase by nearly $100x$ between weeks $23$ and $35$. Write I/O rate, on the other hand, is relatively more stable though it can also change by more than $10x$ over the months. Write I/O appears to fluctuate more in hourly data compared to read I/O. Specifically, while read I/O changes between $1$ and $90$ terabytes, write I/O fluctuates between $0.2$ and $250$ terabytes. We also notice that both read and write operations appear to reach maximum rates around the middle of the day, which can be attributed to increased user interactivity. Similar trends are observed in Theta and Mira clusters, hence we omitted them.

We next look into read and write I/O variations. We first calculated the average ($avg$) and standard deviation ($stdev$) for I/O rates by processing all data points (5-minute interval data). For Blue Waters, average and standard deviation values returned $570$ GiB and $32$ GiB for write operation, $970$ GiB, and $76$ GiB for read I/O, respectively. Next, we count time bins based on their I/O size in comparison to the average in Figure~\ref{fig:Read_Write_STD_above_mean}. We divide each $stdev$ into five sub-intervals and then group them into multiple classes based on the difference between the I/O rate of a time bin and the average I/O rate for the entire dataset. As an example, we place data points whose I/O rates fall between $avg$ and $avg + 0.2\times stdev$ into one group and those whose I/O rates are within $avg + 0.2\times stdev$  and $avg + 0.4\times stdev$ into the next group. As expected, most data points fall within one standard deviation above and below average. We capped the maximum value for the x-axis at $10$  to increase the visibility, but there are few data points in the Blue Waters dataset whose I/O rates are high as $334$ standard deviation above the mean for read and $85$ standard deviation above the mean for write operations. 
\begin{figure*}[t]
\begin{center}
    
\subfigure[Blue Waters]{
    \includegraphics[keepaspectratio=true,angle=0,width=.32\textwidth]{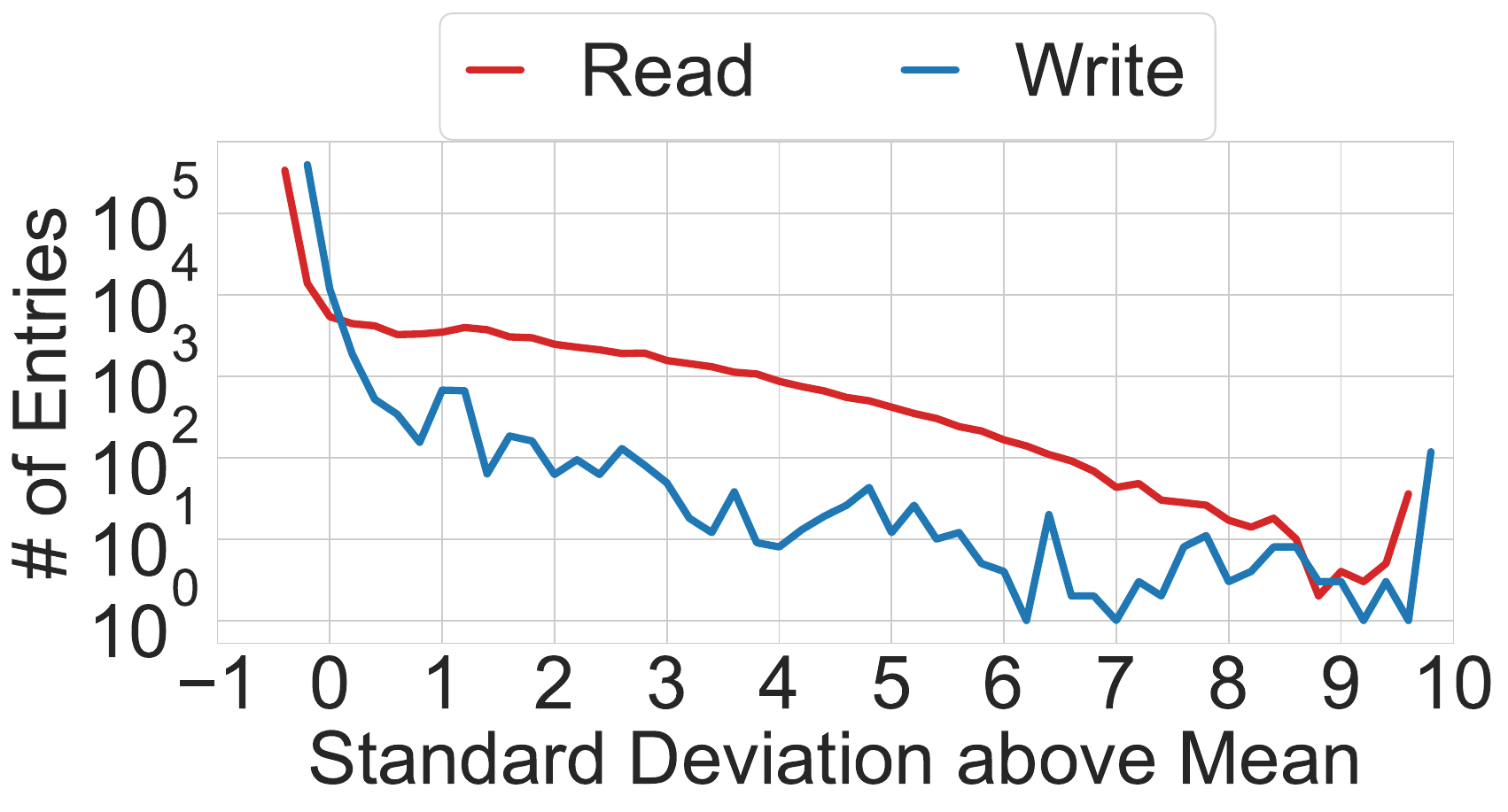}
    \label{fig:Read_Write_STD_above_mean}
}
\hspace{-3mm}
\subfigure[Mira]{
    \includegraphics[keepaspectratio=true,angle=0,width=.32\textwidth]{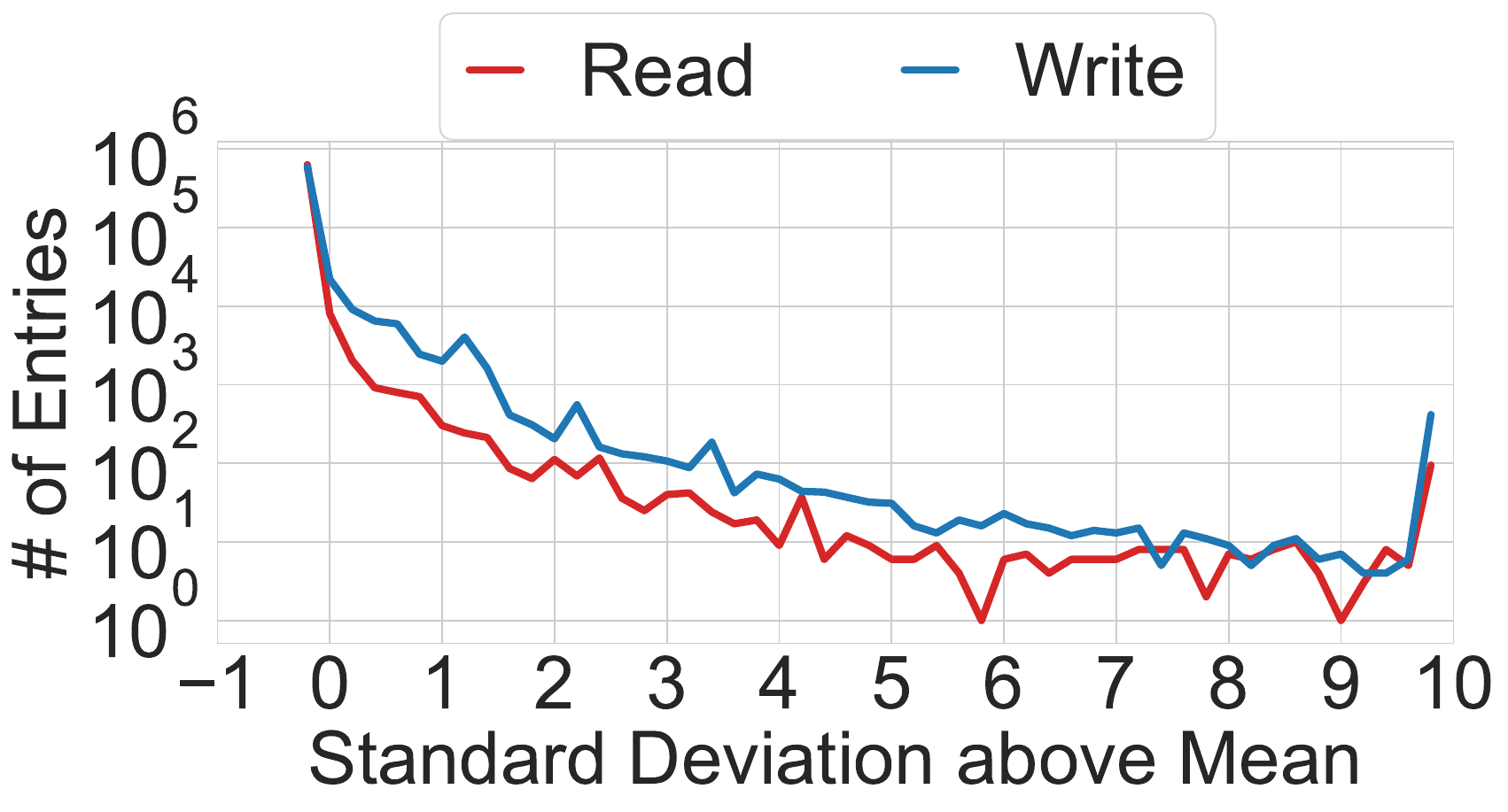}
    \label{fig:Read_Write_STD_above_mean_mira}
}
\hspace{-3mm}
\subfigure[Theta]{
    \includegraphics[keepaspectratio=true,angle=0,width=.32\textwidth]{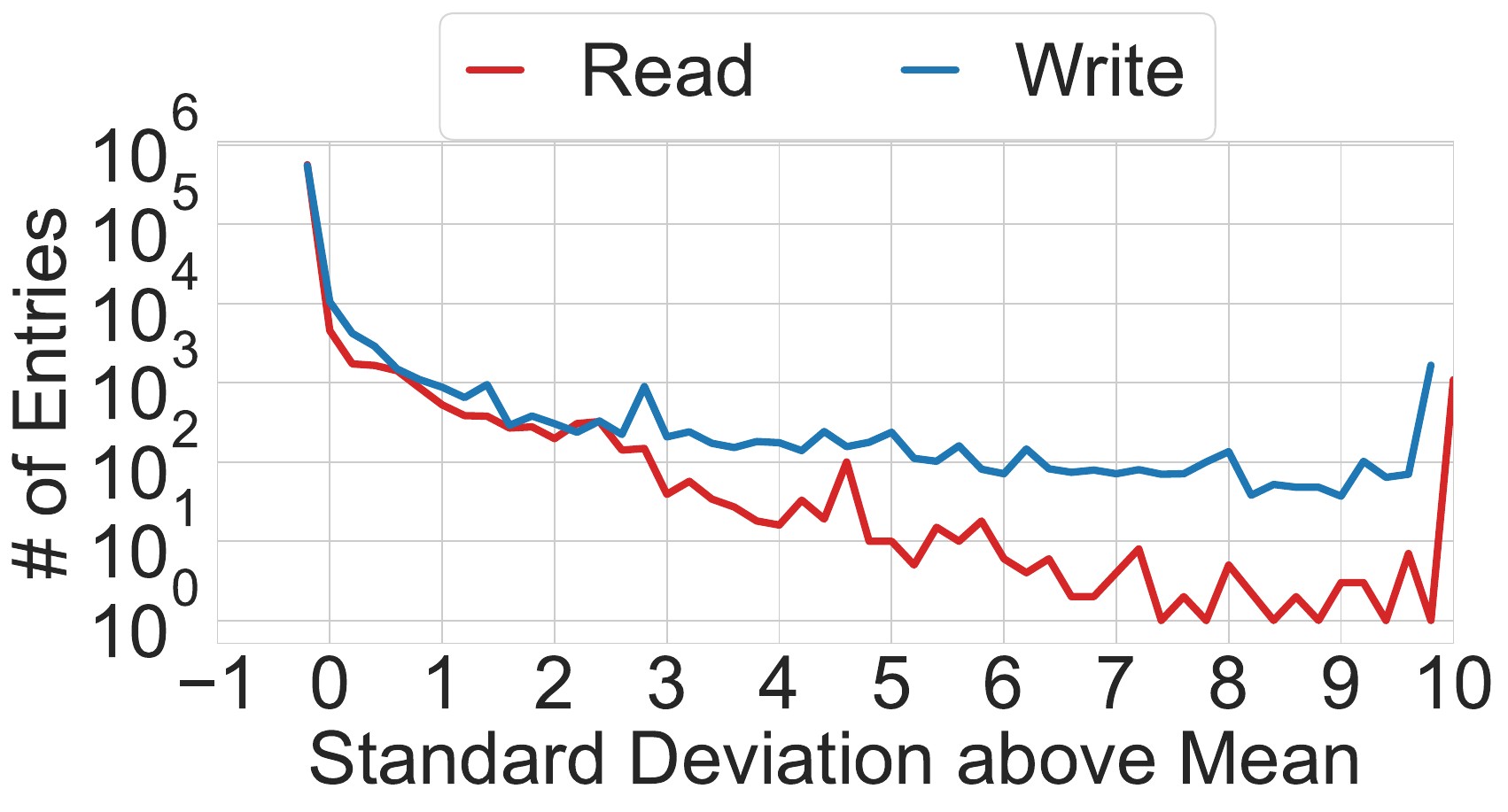}
    \label{fig:Read_Write_STD_above_mean_theta}
}
\hspace{-2mm}
\caption{Variation of I/O rates in comparison to average.}
\vspace{-8mm}
\label{fig:Read_Write_STD_above_mean}
\end{center}
\end{figure*}

Finding a threshold to define a burst is not trivial since selecting a very small threshold will return too many burst intervals, whereas choosing a very large value will not return enough data points to train prediction models. Hence, we selected a threshold value that returns around $1\%$ of all data points (i.e., 5-minute time intervals) as burst. Due to I/O rate differences, this threshold is different for read and write operations. For Blue Waters, we selected $4.5$ standard deviation above the mean for read and $0.5$ standard deviation above the mean for write as burst threshold values. These values became $0.45$ and $1.23$ for Mira and $1.5$ and $4.1$ for Theta. Since I/O operations in the Theta cluster were write-dominated, a larger write threshold value is chosen.

\begin{figure*} [t]
\begin{center}
    \hspace{-3mm}
    \subfigure[Consecutive Read Burst]{
\includegraphics[keepaspectratio=true,angle=0,width=.33\textwidth]{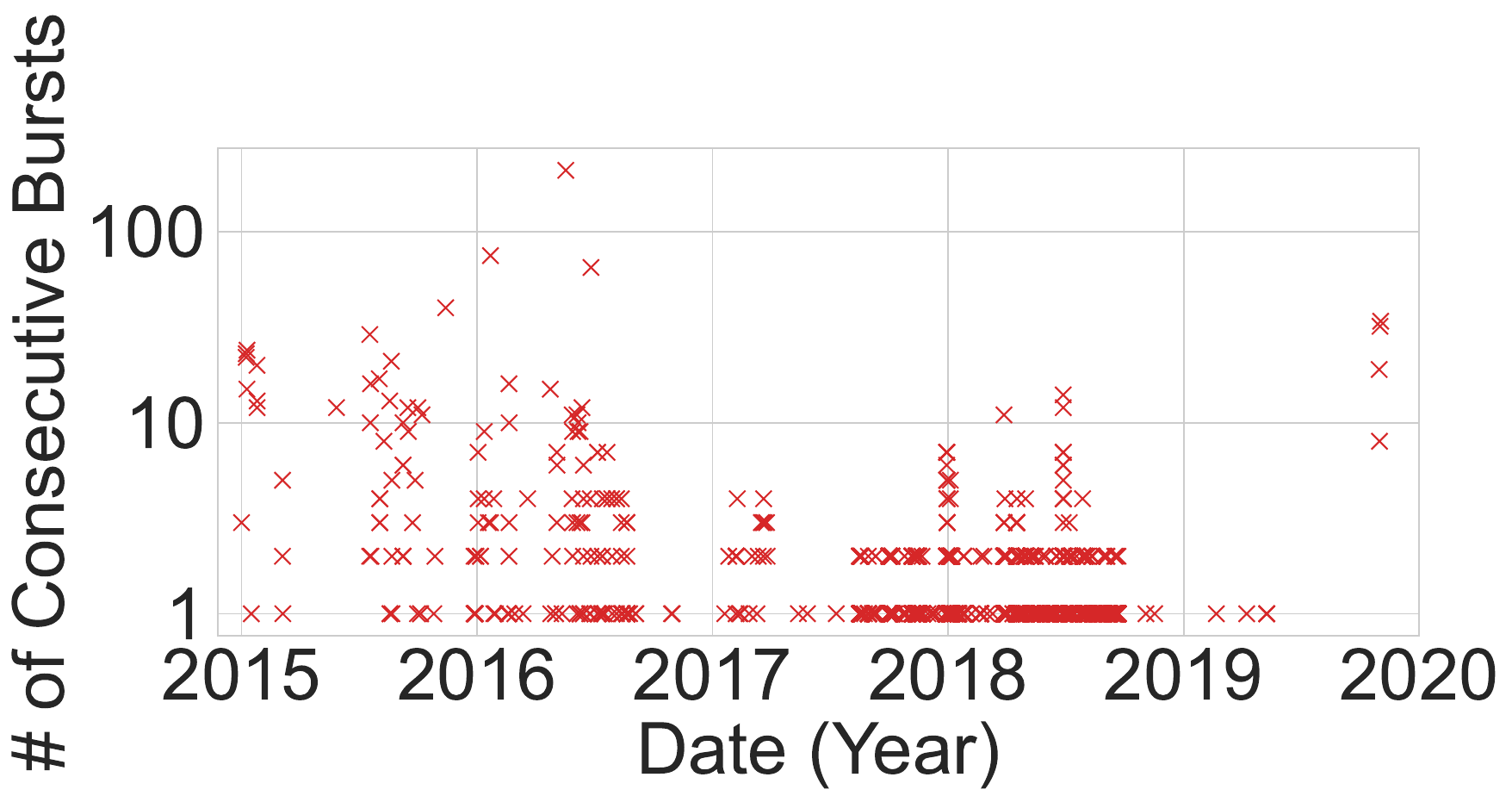}
    \label{fig:Read_bursts_over_time}}
    \subfigure[Consecutive Write Burst]{
\includegraphics[keepaspectratio=true,angle=0,width=.33\textwidth]{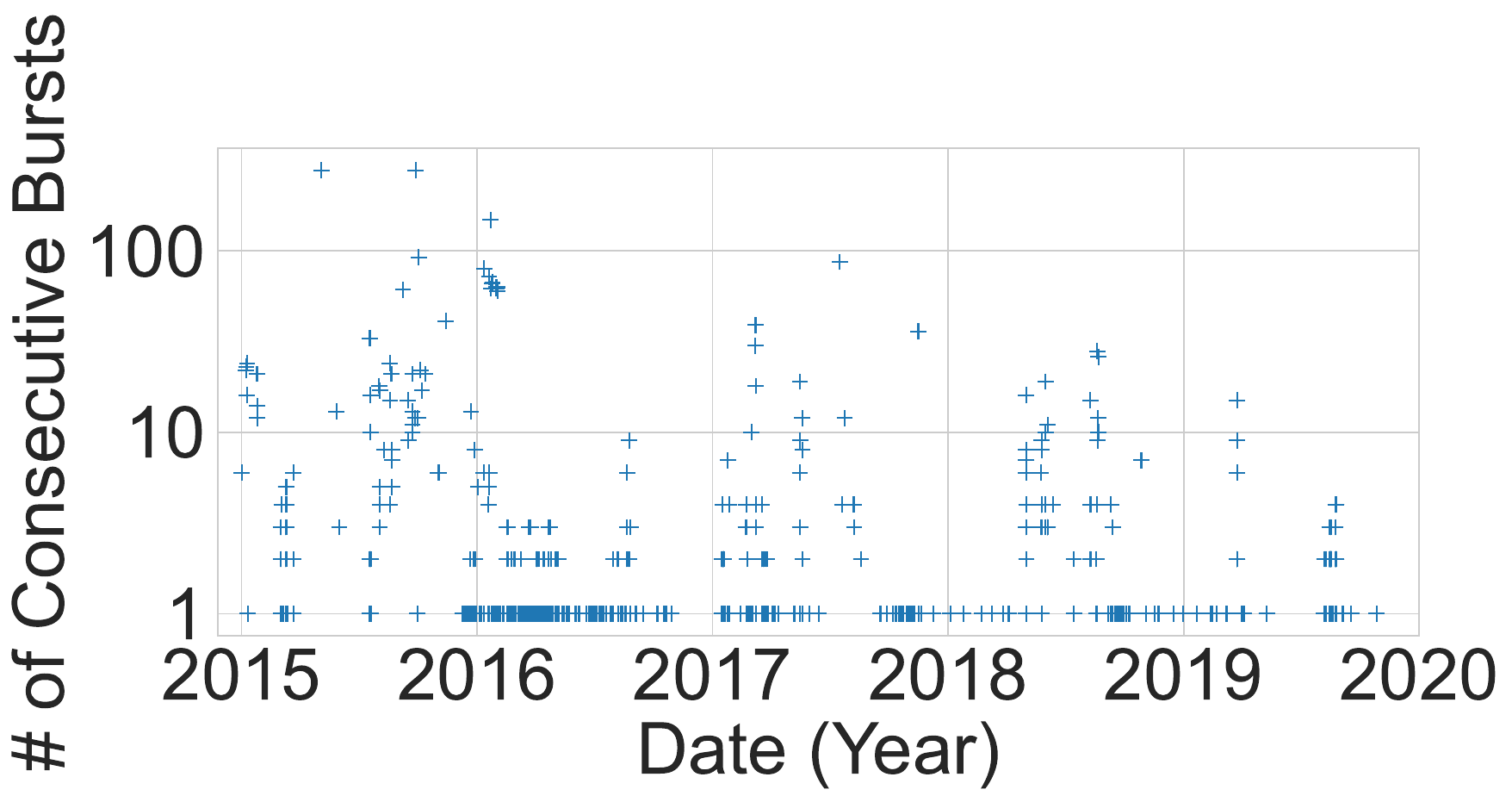}
    \label{fig:Write_bursts_over_time}}
    \hspace{-3mm}
    \subfigure[\# of Consecutive Read Bursts]{
\includegraphics[keepaspectratio=true,angle=0,width=.33\textwidth]{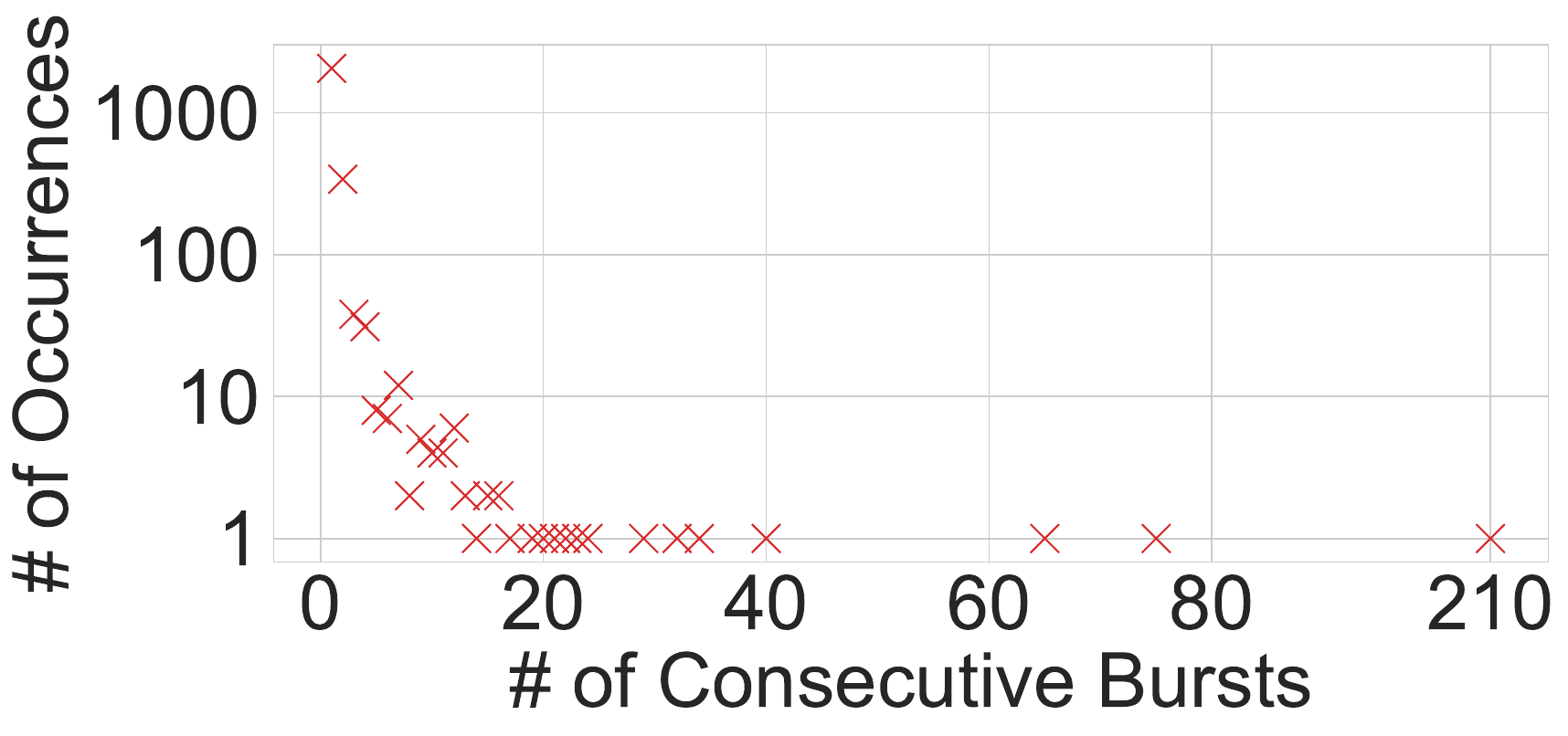}
    \label{fig:cumulative_read_bursts}}
    \hspace{-3mm}
    \subfigure[\# of Consecutive Write Bursts]{
\includegraphics[keepaspectratio=true,angle=0,width=.33\textwidth]{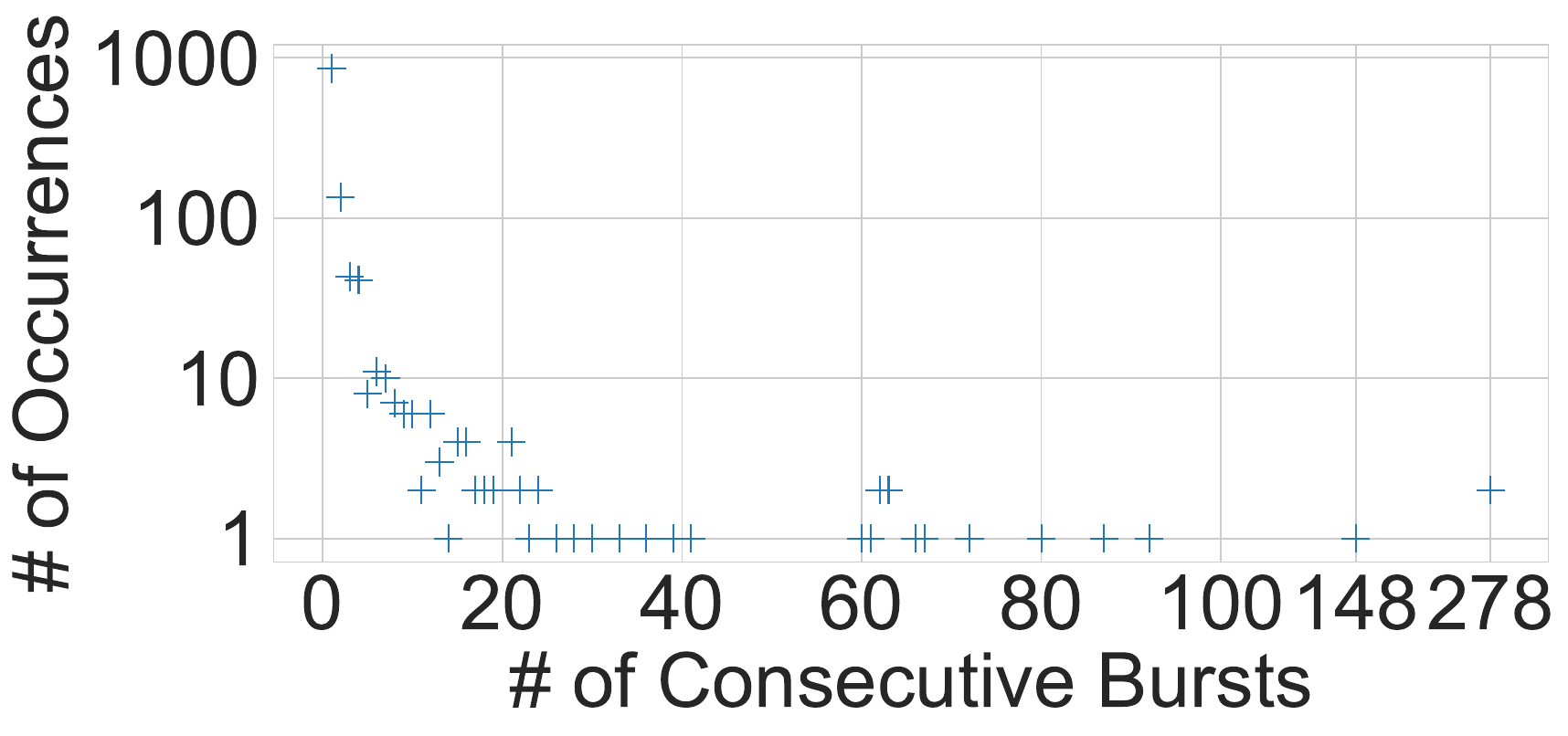}
\label{fig:cumulative_write_bursts}}
\caption{Temporal analysis of read and write I/O bursts. While some bursts occur in consecutive time intervals, most bursts last only one or two time intervals, turning burst prediction into a nontrivial task.}
\vspace{-5mm}
\label{fig:Bursts_over_time}
\end{center}
\end{figure*}

As we aim to predict the bursts before they happen, we next look into the temporal relationship between the bursts. In other words, we measure how often bursts appear next to each other, which can be used to understand the difficulty of the prediction task. For example, if most bursts happen consecutively, one can achieve a high-accuracy prediction simply by copying the current timestep's burst state (i.e., a burst of non-burst). Figure~\ref{fig:Bursts_over_time} presents the temporal relationship between bursts for read and write operations for the Blue Waters dataset. Figure~\ref{fig:Read_bursts_over_time} and~\ref{fig:Write_bursts_over_time} show the number of consecutive burst occurrences over the observation period. It is clear that more than half of all bursts are not followed by another burst interval, and more than $90\%$ of bursts appear next to $10$ or fewer other bursts. Figure~\ref{fig:cumulative_read_bursts} and~\ref{fig:cumulative_write_bursts} show the number of times consecutive bursts happened. We can again see that a significant portion of all bursts (around $1,000$, which corresponds to $25\%$ of all bursts) does not come before or after another burst; i.e., the number of consecutive bursts is equal to one. Thus, it is not trivial to predict bursts with high accuracy since the majority of them appear after a non-burst time interval. Therefore, we apply machine learning models to forecast read and write I/O bursts. 

\section{I/O Burst Prediction}
To train machine learning models, we identified three sets of parameters. The features in the first set are the number of read/write bytes of a given interval; hence the prediction is made based on the I/O rate of the previous time interval. This feature set represents a simple solution that only uses one feature (read and write prediction models are trained separately, thus each model only uses one feature). The second set contains three features; read/write bytes, the number of read/write operations, and the total time spent on read/write operations. Since some applications may experience I/O contention even if the file system is not fully utilized due to multiple applications accessing the same storage server simultaneously, this feature set includes the total I/O time that HPC applications spent during a measurement interval. The third feature set includes all the features used in the second feature set and incorporates a few additional features to capture the temporal nature of the data. Specifically, we added Exponential Moving Average (EMA) and Moving Average Convergence Divergence (MACD) for read/write bytes to keep track of I/O behavior over longer periods. For example, if an application issues heavy I/O requests periodically, incorporating more historical data will allow better prediction performance.

\begin{figure}[t]
\begin{center}
    \hspace{-3mm}
    \subfigure[Read Burst Prediction]{
    \includegraphics[keepaspectratio=true,angle=0,width=.24\textwidth]{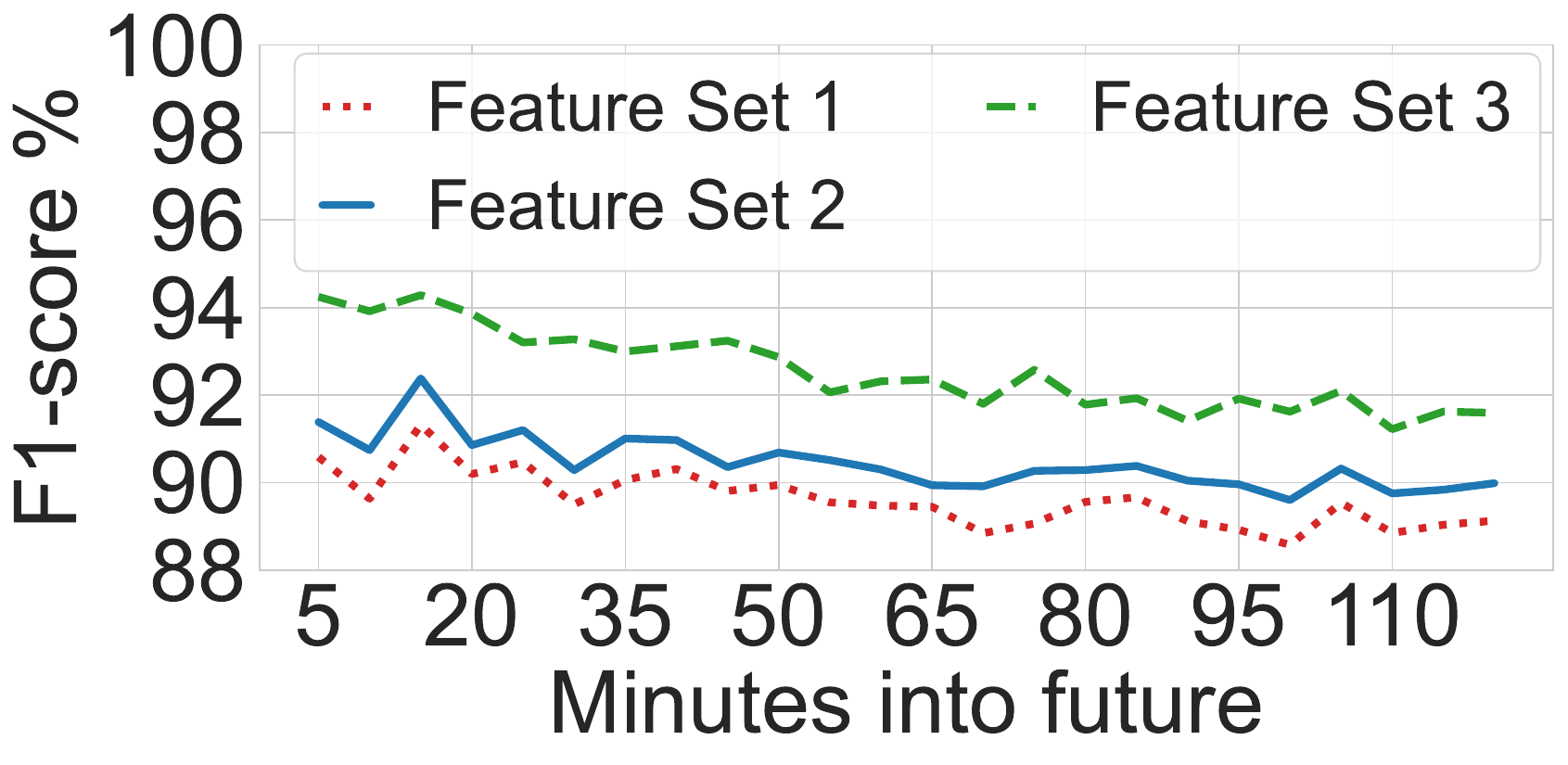}
    \label{fig:RF_read_burst_acc_over_time}}
    \hspace{-4mm}
    \subfigure[Write Burst Prediction]{
\includegraphics[keepaspectratio=true,angle=0,width=.24\textwidth]{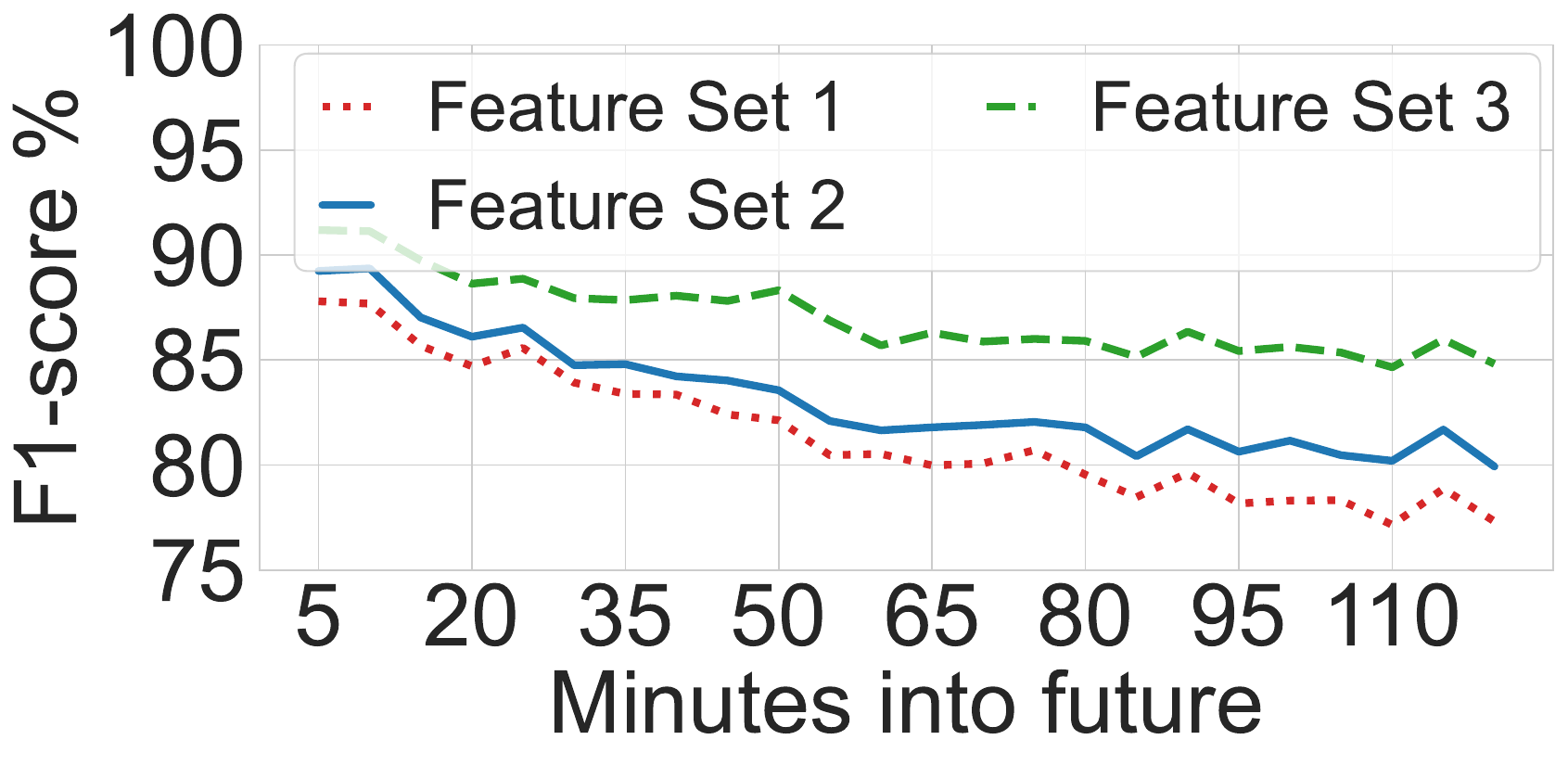}
    \label{fig:RF_write_burst_acc_over_time}}
    \vspace{-3mm}
\caption{Comparison of Random Forest-based I/O burst prediction using different sets of features. Feature set 3 incorporates the temporal behavior of I/O size.}
\vspace{-8mm}
\label{fig:RF_acc_over_time}
\end{center}
\end{figure}

EMA is a type of moving average that puts more weight on recent data. It is calculated using Equation~\ref{eq:eq1} and~\ref{eq:eq2}, where $current$ is the last observed value of a feature. The value of $\alpha$ affects how fast the new data is adopted, and $n$ is the window size (i.e., timesteps). For example, an EMA that covers $10$ timesteps will adapt to new values much more quickly than one that covers $100$ timesteps; thus, it gives higher weight to recent data. MACD was designed to reveal the changes in the strength of a trend~\cite{MACD}. It is used with two EMAs that cover different time frames, as given in Equation~\ref{eq:eq3}. Since short-range EMAs are more sensitive to recent data and vice versa for long ones, it is possible to gauge current trends based on whether the result is positive or negative. We created two MACDs using four different EMAs. The first uses a 60-minute EMA and a 130-minute EMA to capture shorter-term trends. Since our data set uses five-minute intervals, this corresponds to alpha values of $2/13$ and $2/27$, respectively. The second one uses a 500-minute EMA and a 1000-minute EMA to capture medium-term trends. Unlike first and second feature sets, these MACDs can provide significantly more information on I/O trends that span across days.

\begin{equation} \label{eq:eq1}
    E_t = \alpha \times current + (1-\alpha) \times E_{t-1}
\end{equation}
\begin{equation} \label{eq:eq2}
    \alpha = \frac{2}{n + 1}
\end{equation}
\begin{equation} \label{eq:eq3}
    MACD = E_t - E_{t-1}
\end{equation}

We trained Random Forest models with all three sets of features using the Blue Waters dataset. Although data points are originally binned into 5-minute intervals, we created $20$ time intervals by combining data from multiple consecutive 5-minute intervals. We hence developed a separate model for each time interval to evaluate the accuracy of models when estimating burst predictions up to two hours ahead of time. 
Our choice of a two-hour window for the job’s runtime is based on previous research which showed that the median job runtime on Titan and Cori supercomputers are $1.58$ and $1.95$ hours respectively~\cite{wan2020performance}. Similarly, Paul et al. analyzed job statistics collected from Cab and Quartz Clusters and showed that more than $90\%$ of jobs
run for less than 2 hours \cite{paul2020understanding}. Figure~\ref{fig:RF_acc_over_time} shows the accuracy (F-score) of the trained Random Forest models.  We tested the models for read (Figure~\ref{fig:RF_read_burst_acc_over_time}) and write (Figure~\ref{fig:RF_write_burst_acc_over_time}) separately. Each model was trained using the first $80$\% of time periods and tested using the last $20$\% of time periods. We observe that the third feature set attains the best overall performance, which indicates the benefit of incorporating historical data (i.e., EMA and MACD). 

We next compared Logistic Regression, Gaussian Naive Bayes, Decision Tree, Random Forest, and XGBoost models. We leverage scikit-learn library to tune the hyperparameters of the selected models. Specifically, we tuned tree depth, maximum leaf node,  and number of features for splitting for Decision Tree, the number of trees and tree depth, maximum leaf node, number of estimators for Random Forest, booster, max depth, minimum child weight, gamma, regularization parameter (alpha and lambda), and colsample\_bytree ratio for XGBoost, variance smoothing value for Gaussian Naive Bayes, and penalty, solver, inverse of regularization strength, and maximum number of iterations for the Logistic Regression.
 We used the saga solver for the Logistic Regression model and set max iterations to $15,000$ to ensure it does not terminate before converging. We default hyperparameters for the Naive Bayes model. We present performance results using the F-1 score, which is the harmonic mean of precision and recall metrics. Precision is the portion of the true positive ($TP$) predictions over the total number of positive predictions (i.e., $TP+ FP$)). In this study, the true positive is the amount of I/O bursts that are truly predicted. The false positive, on the other hand, is an incorrect prediction of burst occurrences.

\begin{figure*}[t]
\begin{center}
    \subfigure[Blue Waters- Read]{\includegraphics[keepaspectratio=true,angle=0,width=.32\textwidth]{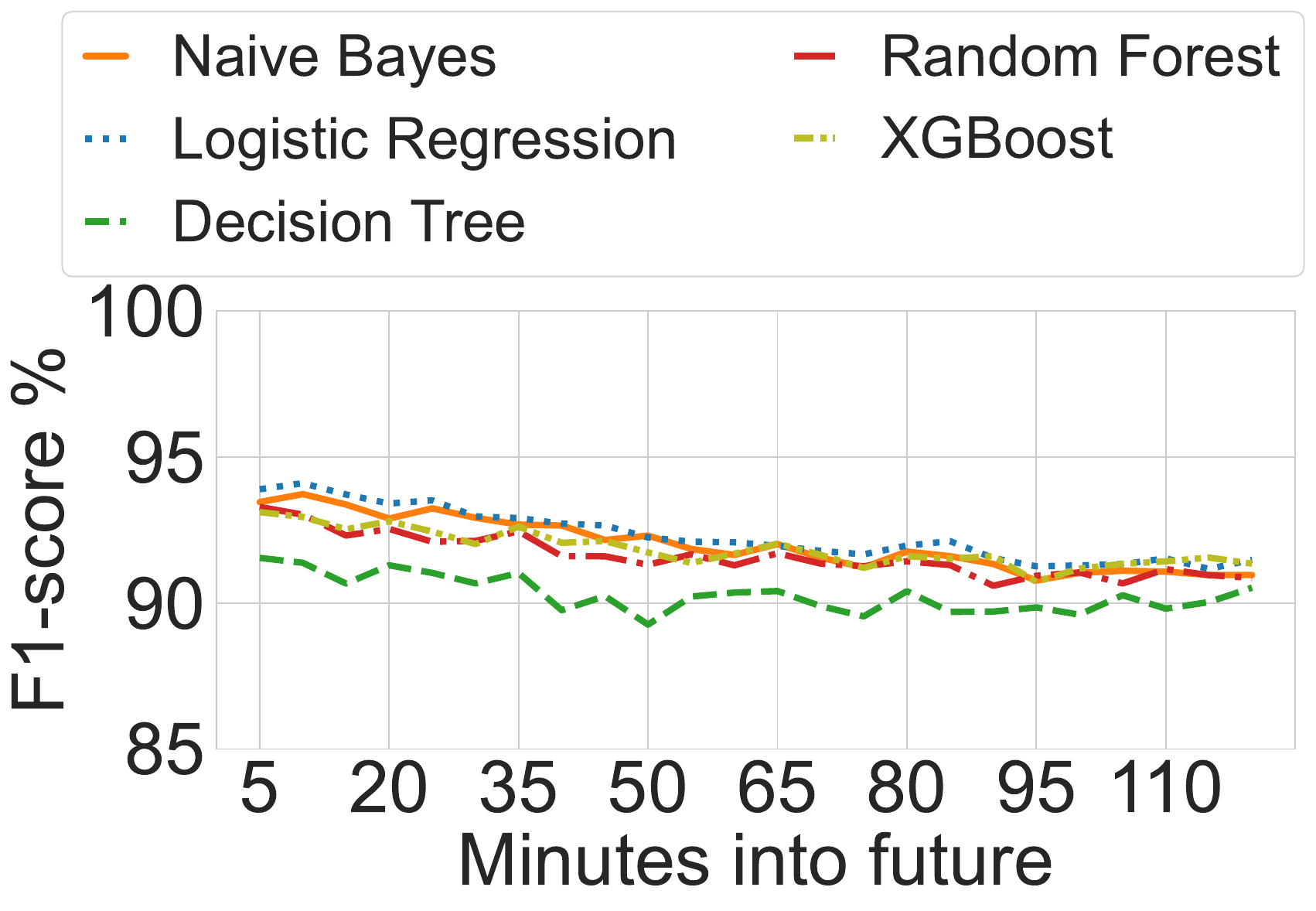}
    \label{fig:Models_Read_F1_over_time_2}}
    \hspace{-4mm}
    \subfigure[Blue Waters- Write]{\includegraphics[keepaspectratio=true,angle=0,width=.32\textwidth]{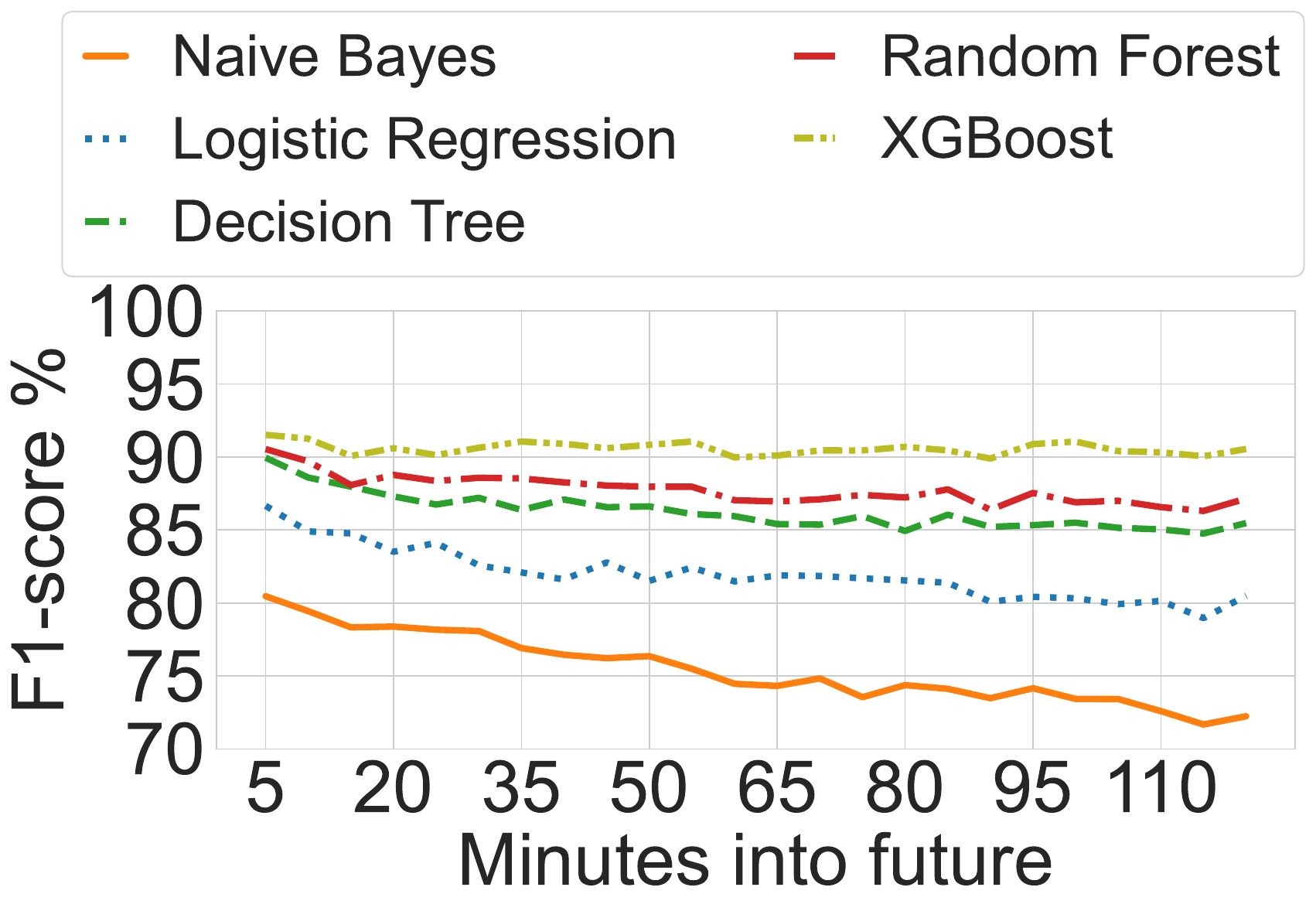}
    \label{fig:Models_Write_F1_acc_over_time}}
    \hspace{-4mm}
    \subfigure[Mira- Read]{\includegraphics[keepaspectratio=true,angle=0,width=.32\textwidth]{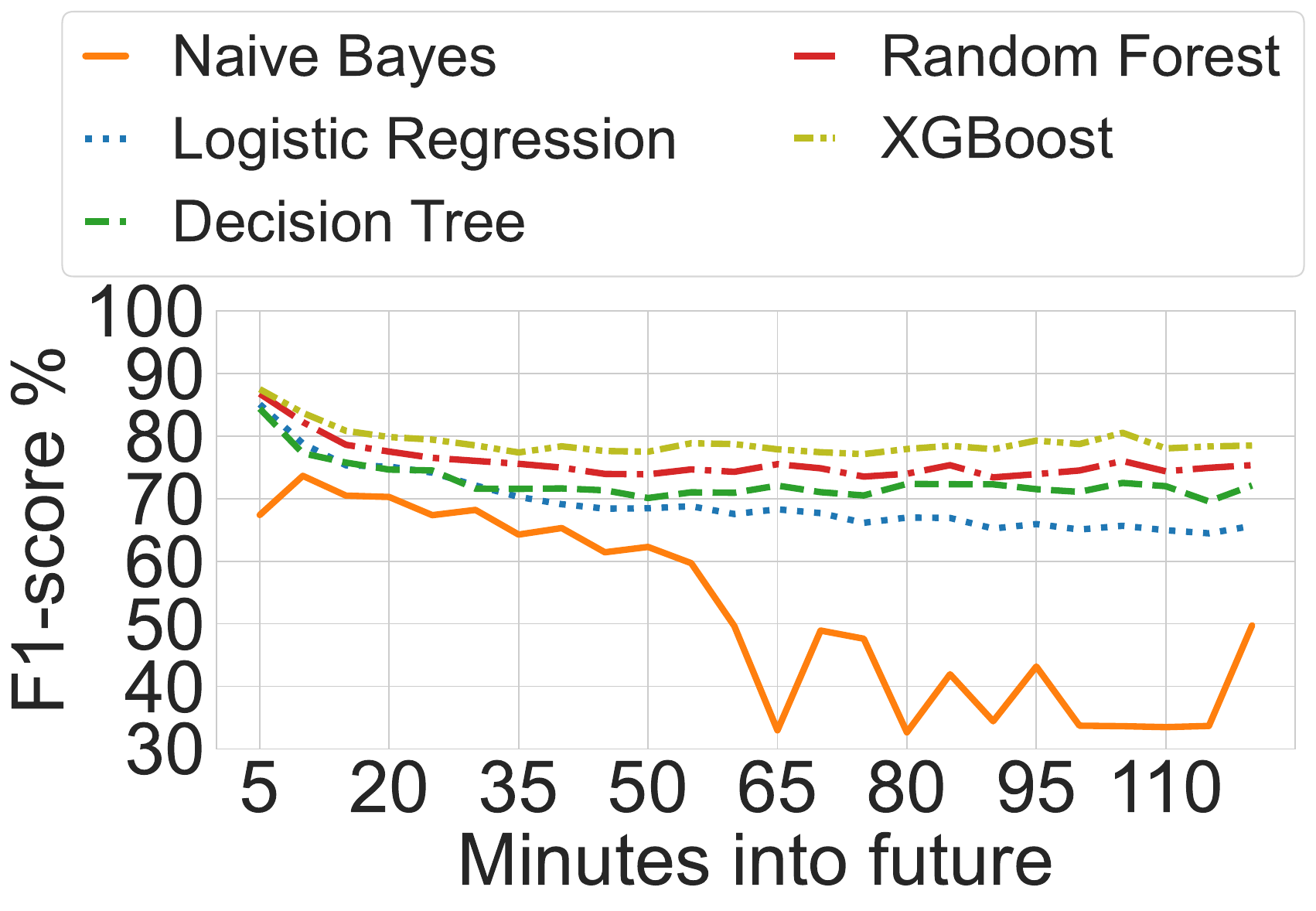}
    \label{fig:Models_Read_F1_over_time_mira}}
    \hspace{-4mm}
    \subfigure[Mira- Write]{\includegraphics[keepaspectratio=true,angle=0,width=.32\textwidth]{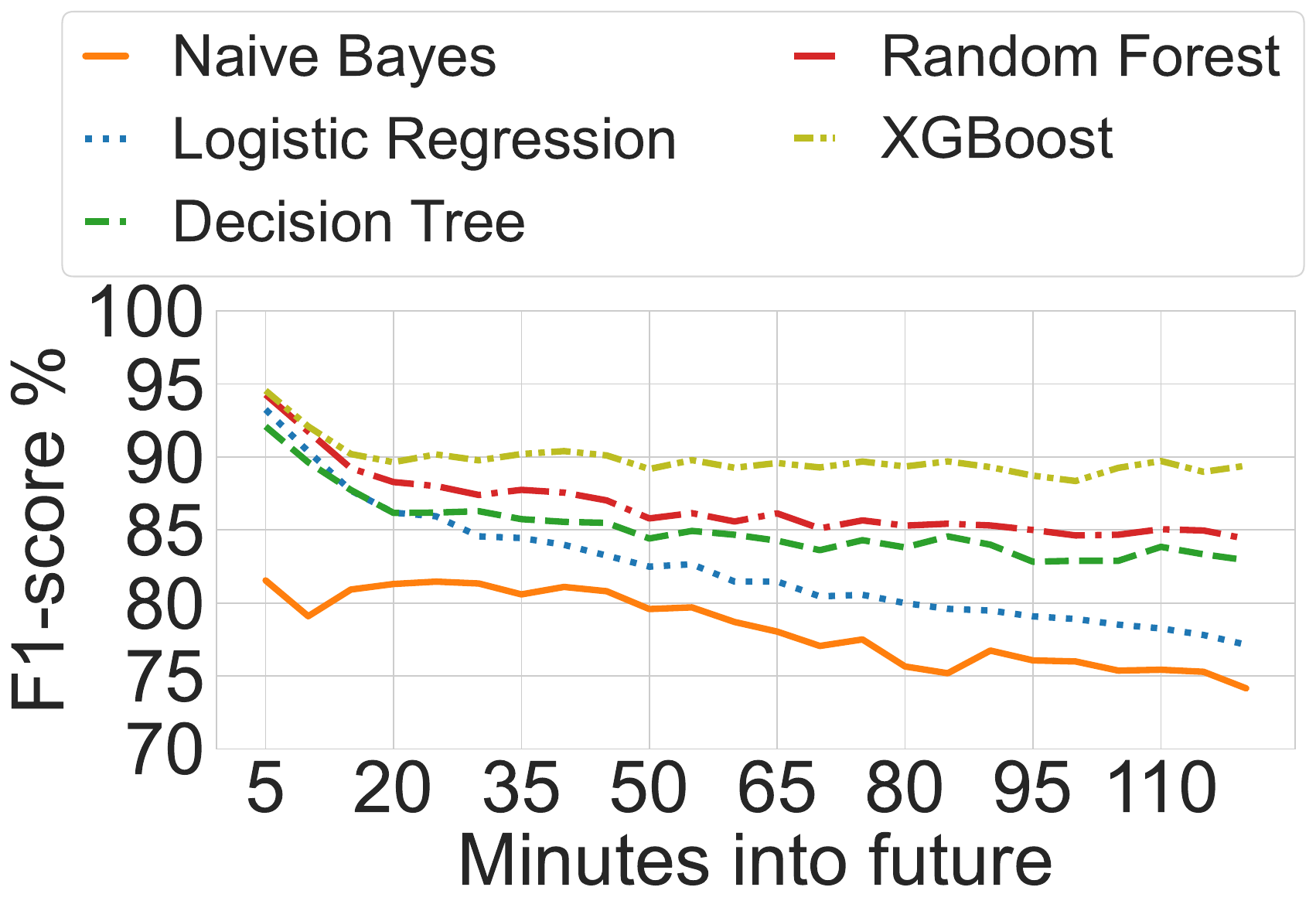}
    \label{fig:Models_Write_F1_over_time_mira}}
    \hspace{-4mm}
    \subfigure[Theta- Read]{\includegraphics[keepaspectratio=true,angle=0,width=.32\textwidth]{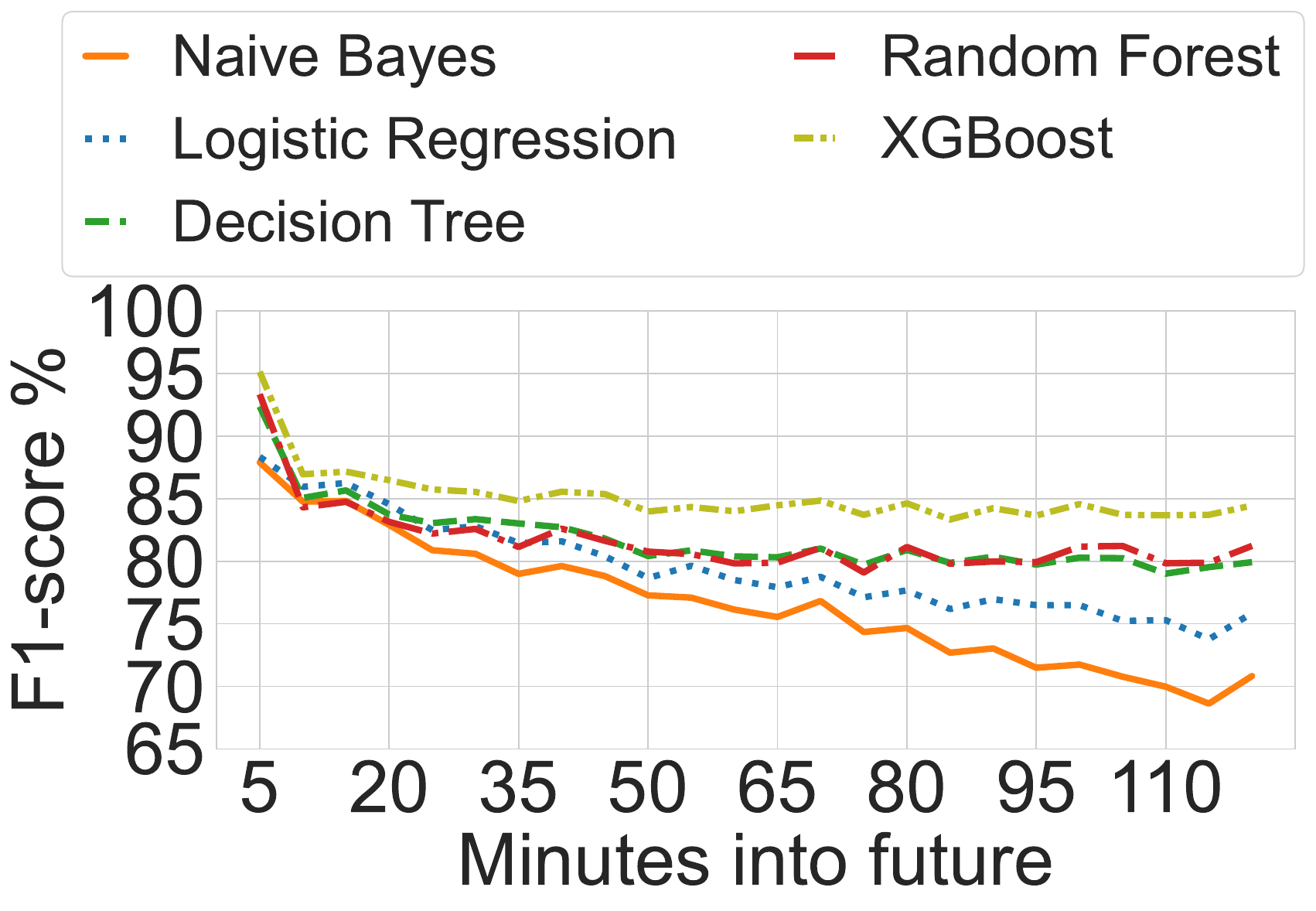}
    \label{fig:Models_Read_F1_over_time_theta}}
    \hspace{-4mm}
    \subfigure[Theta- Write]{\includegraphics[keepaspectratio=true,angle=0,width=.32\textwidth]{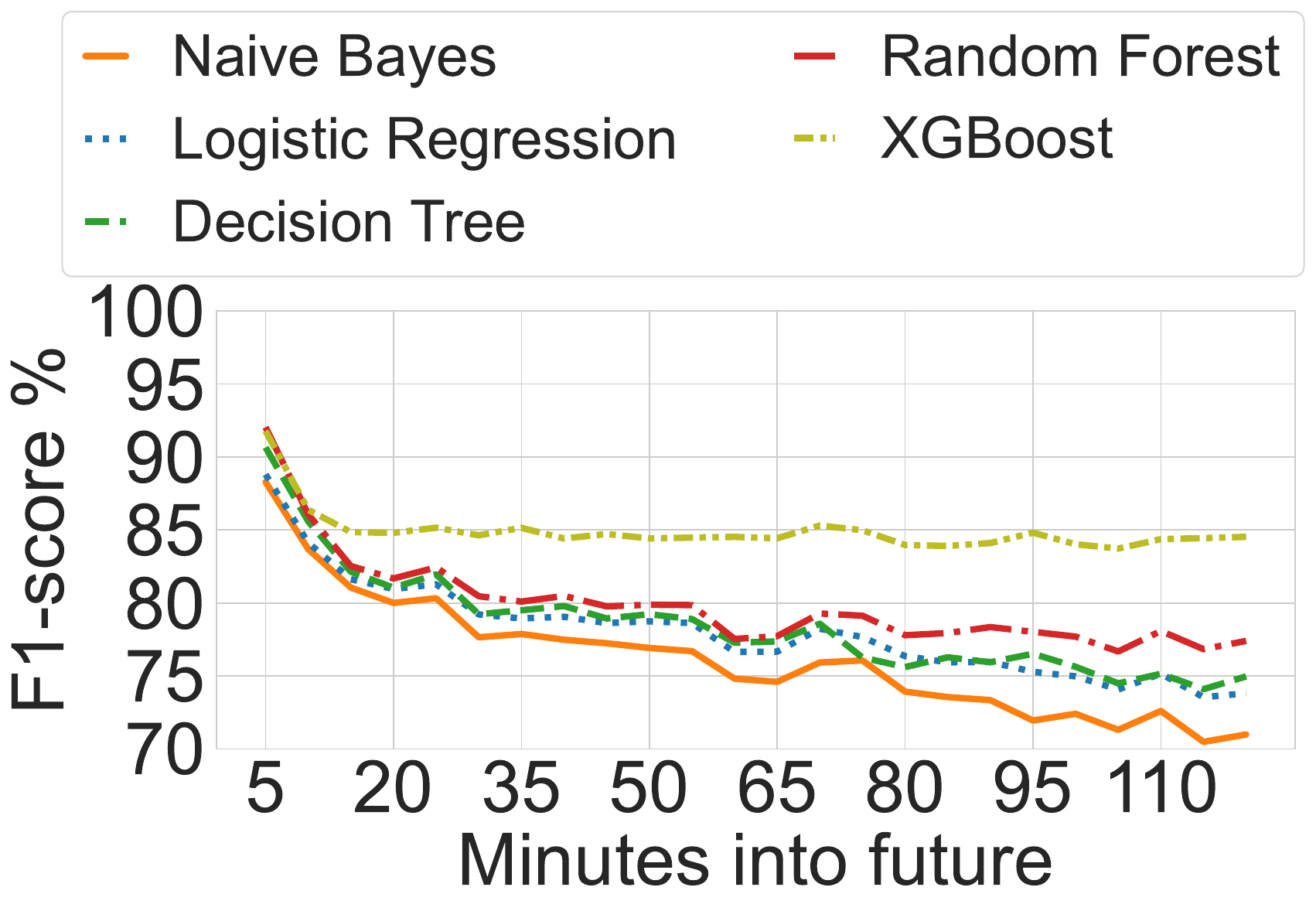}
    \label{fig:Models_Write_F1_over_time_theta}}
    \vspace{-2mm}
\caption{Performance of different machine learning models for read and write burst prediction. XGBoost attains the best overall performance with nearly $90\%$ score both for read and write I/O predictions. Its read burst performance degrades $2-3\%$ when prediction time is increased from 5 minutes to 120 minutes. }
\label{fig:Models_evaluation_over_time}
\end{center}
\end{figure*}

Figures \ref{fig:Models_Read_F1_over_time_2} and \ref{fig:Models_Write_F1_acc_over_time} compare the performance of the ML models for the Blue Waters dataset. We observe that although the prediction models are more accurate for shorter time frames, their performance is only slightly degraded for longer prediction windows. As an example, the prediction accuracy for Logistic Regression is decreased from $94\%$ to $92\%$ for read burst prediction when the prediction interval is increased from $5$ minutes to $120$ minutes. Overall, we find that the XGBoost model attains the best performance by yielding more than $90\%$ accuracy for both read and write I/O burst prediction with any time interval. We believe that this level of accuracy is likely sufficient to develop preventive measures such as coordinating I/O operations and putting non-essential applications (e.g., file system scrubbers) on hold.


Mira and Theta cluster results follow a similar pattern with Blue Waters; a slight decrease in model performance as the size of time bin size is increased from 5-minute to 120-minute. The XGBoost model again outperforms the other models as its performance is affected the least by the increased time interval. However, the models achieved lower F1 scores than the Blue Waters dataset which can be attributed to the fact that Darshan logs from Mira and Theta supercomputers are coarser-grained compared to the Blue Waters dataset. In other words, Darshan logs from Blue Waters report at file-level precision for each process  whereas they are only process-level for Mira and Theta clusters as discussed in Section 2.

\newcolumntype{Q}[1]{>{\centering\arraybackslash}p{#1}}
\begin{table*}[t]
\centering
\caption{F1-score for multi-level I/O burst prediction.}
\label{table:2}
\begin{tabular}{Q{3.5cm} | Q{1.3cm}  Q{1.3cm} | Q{1.3cm}  Q{1.3cm} | Q{1.3cm}  Q{1.3cm}} 
\bf  & \multicolumn{2}{c}{\bf Blue Waters} & \multicolumn{2}{c}{\bf Mira} & \multicolumn{2}{c}{\bf Theta} \\
\hline
\bf ML Model &  \bf Read &\bf Write  & \bf Read &\bf Write & \bf Read & \bf Write \\
\hline
 Naive Bayes &  71.60 & 79.44  & 69.69  & 81.67 & 80.50 & 61.54 \\
 Logistic Regression & 71.78  & 74.53  & 79.21 & 81.3 & 76.09 & 61.62 \\
 Decision Tree &  71.57 & 85.26  & 79.97 & 83.49 & 88.91 & 65.16 \\
 Random Forest & 72.64  &  86.81  & 83.20 & 86.36 & 91.01 & 69.41 \\
 XGBoost &  72.20 & 87.24  & 83.47 & 86.33 & 91.66 & 70.87 \\
\hline
\end{tabular}
\vspace{-5mm}
\end{table*}

\subsection{Predicting the Severity of I/O Bursts}
A non-negligible number of I/O bursts have an I/O rate that is more than $10$ standard deviation above the mean in all clusters as shown in Figure~\ref{fig:Read_Write_STD_above_mean}. Hence, using a single cutoff point to define I/O burst can be too restrictive as the degree of the burst can affect the possible action delaying some I/O-intensive jobs altogether if the system experiences a significant burst. Hence, we next look at splitting burst classes into multiple groups to differentiate the severity of bursts using 5-minute time bin data. To do so, we divide the burst data points into $5$ subgroups based on I/O rates. Specifically, we split the interval between the original burst cutoff points and the $10$ standard deviation above the mean into five equal sizes. For example, $1.23$ above the mean is used to define burst for write operation in Mira. To define the burst levels, we create five equal range cutoff points between $1.23$ and $10$ standard deviation above the mean. Bursts that are more than $10$ standard deviation are placed into the last burst group. This turns the prediction from binary to multi-label classification with six classes (five bursts and one non-burst). As expected, the overall performance decreased compared to the burst/no-burst binary classification. This decline in performance can be attributed to the fact that only a small percentage of the data is labeled as bursts; thus dividing this data into multiple ranges could have impacted the model's performance due to a lack of sufficient training data. Regardless, XGBoost and Random Forest still achieve considerably higher performance than others with more than $70\%$ accuracy, as shown in Table~\ref{table:2}.

\begin{table}[]
\centering
\caption{Hardware Details}
\begin{tabular} {Q{2cm} | Q{5cm}}
\hline
\textbf{Component} & \textbf{Specification} \\ \hline
CPU & Intel i5-6600 CPU @ 3.30GHz \\ \hline
RAM & 16GB DDR4 @ 2400MHz \\ \hline
Storage & 256GB SATA SSD \\ \hline
\end{tabular}
\label{tab:hardware-details}
\end{table}
\vspace{-3mm}

\subsection{Burst-Aware Job Scheduling}
We develop a burst-aware job scheduling algorithm to demonstrate the benefits of the system-level write I/O burst prediction on application performance by avoiding I/O bursts. We utilize the Deep Learning Input/Output (DLIO) benchmark \cite{devarajan2021dlio} to evaluate the performance of the proposed scheduling algorithm. The DLIO benchmark simulates deep-learning applications' input/output (I/O) characteristics. In our experiments, we followed the example available at \cite{dlio_benchmark_examples} to emulate the behavior of three-dimensional semantic segmentation of brain tumors (UNET3D) using convolutional neural network~\cite{UNET3D}. We configured the total number of epochs to five, the model size to 3GB, and checkpointing frequency to once in every epoch. We run the benchmark on a server whose hardware specifications are listed in Table \ref{tab:hardware-details}. To generate system-level write bursts, we execute a background task that writes to the file system at the maximum possible rate while the UNET3D application is running. Since we make burst predictions in five minutes intervals, the background task is controlled at five minutes frequency; i.e., it either runs for the entire five-minute interval (if a burst is predicted) or does not run at all. 
\begin{algorithm}
\caption{I/O Burst Aware Scheduling Algorithm}\label{alg:burst-awarea-sched}
\begin{algorithmic}[1]
\Require{\text{$burst\_pattern$, $min\_time$, $max\_time$,} \text{$burst\_length$,} $\alpha$}
\Ensure{$output$: $start\_time$}
    \State $schedule\_options = \{\}$
    \State \text{$progress_{nb} =$ $\frac{min\_time}{\lceil \frac{min\_time}{burst\_length} \rceil}$} \Comment{Progress w/o burst} \label{line:progress_nb}
    \State \text{$progress_b =$ $\frac{min\_time}{\lceil \frac{max\_time}{burst\_length} \rceil}$} \Comment{Progress w/ burst}\label{line:progress_b}
    
    
    
    \State $best\_score = \infty$
    \For{$i \gets 1$ to $N$}\Comment{Iterate over 5-min intervals, N=24}  \label{line:iterate}
        
        \State \text{$total\_work =min\_time$}\label{line:time_estimate_start}
        \State \text{$remaining\_work = total\_work$}
        \State $j=i$
            \While{$remaining\_work >0 $}
                    \If{$burst\_pattern[j]$ == 1} \Comment{Burst case}
                        \State $remaining\_work -= progress_b$
                    \Else
                        \State $remaining\_work -= progress_{nb}$
                    \EndIf
                \State $j \gets j + 1$ 
            \EndWhile
            \State $delay = i * pred\_interval$  \label{line:time_estimate_end}
            \State $run\_time = (j-i) * pred\_interval$
            \State $score = \alpha \times delay + (1-\alpha) \times run\_time$
            \If{$score < best\_score$}
                \State $best\_score = score$
                 \State $start\_time = i * pred\_interval$
            \EndIf
        
    \EndFor

    
    

    \State \textbf{return} $start\_time$
\end{algorithmic}
\end{algorithm}

\begin{figure}[]
\begin{center}
\includegraphics[keepaspectratio=true,angle=0,width=.38\textwidth]{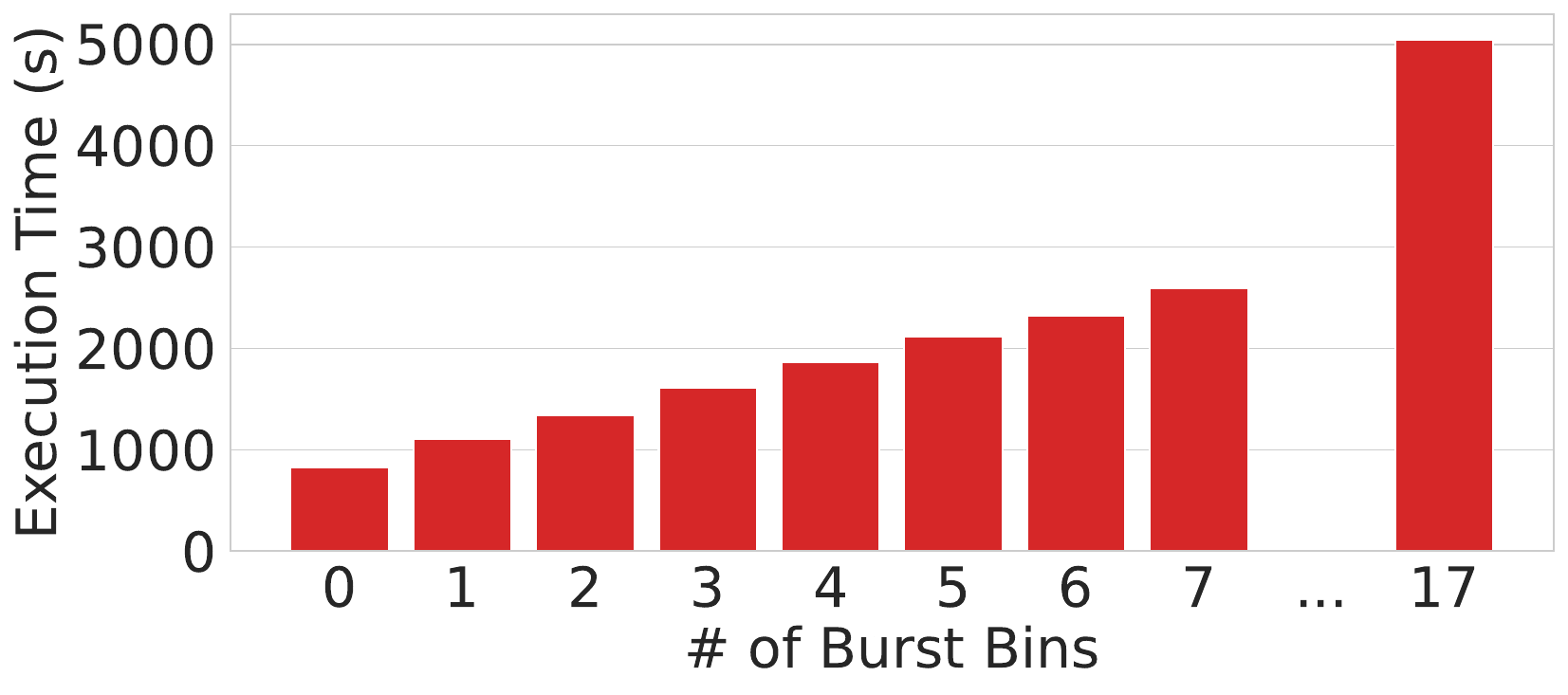}

\caption{The impact of I/O burst on the execution time of UNET3D job. While the execution time is around $800$ seconds when executed without I/O bursts, it increases by around $30\%$ for every overlapping burst bin (each bin is five-minute long) and reaches $5,000$ seconds when executed under bursty I/O completely.}
\vspace{-8mm}
\label{fig:Executiontime_vs_burst_length}
\end{center}
\end{figure}

Figure \ref{fig:Executiontime_vs_burst_length} illustrates the impact of write I/O burst on application execution time. In the absence of bursts, the application finishes in around $850$ seconds. As the number of time bins the application's execution coincides with a burst (i.e., running along with the background task) increases, the execution time increases considerably. As an example, if the application is exposed to burst in the first five minutes of its execution (I/O burst interval is $1$), then the execution time increases by nearly $30\%$ and becomes $1,020$ seconds. If it runs in the presence of I/O bursts all the time, the execution time increases by more than five times.  


\begin{figure*}
\begin{center}
    \subfigure[Minimum Delay Time (MDT)]{\includegraphics[keepaspectratio=true,angle=0,width=.32\textwidth]{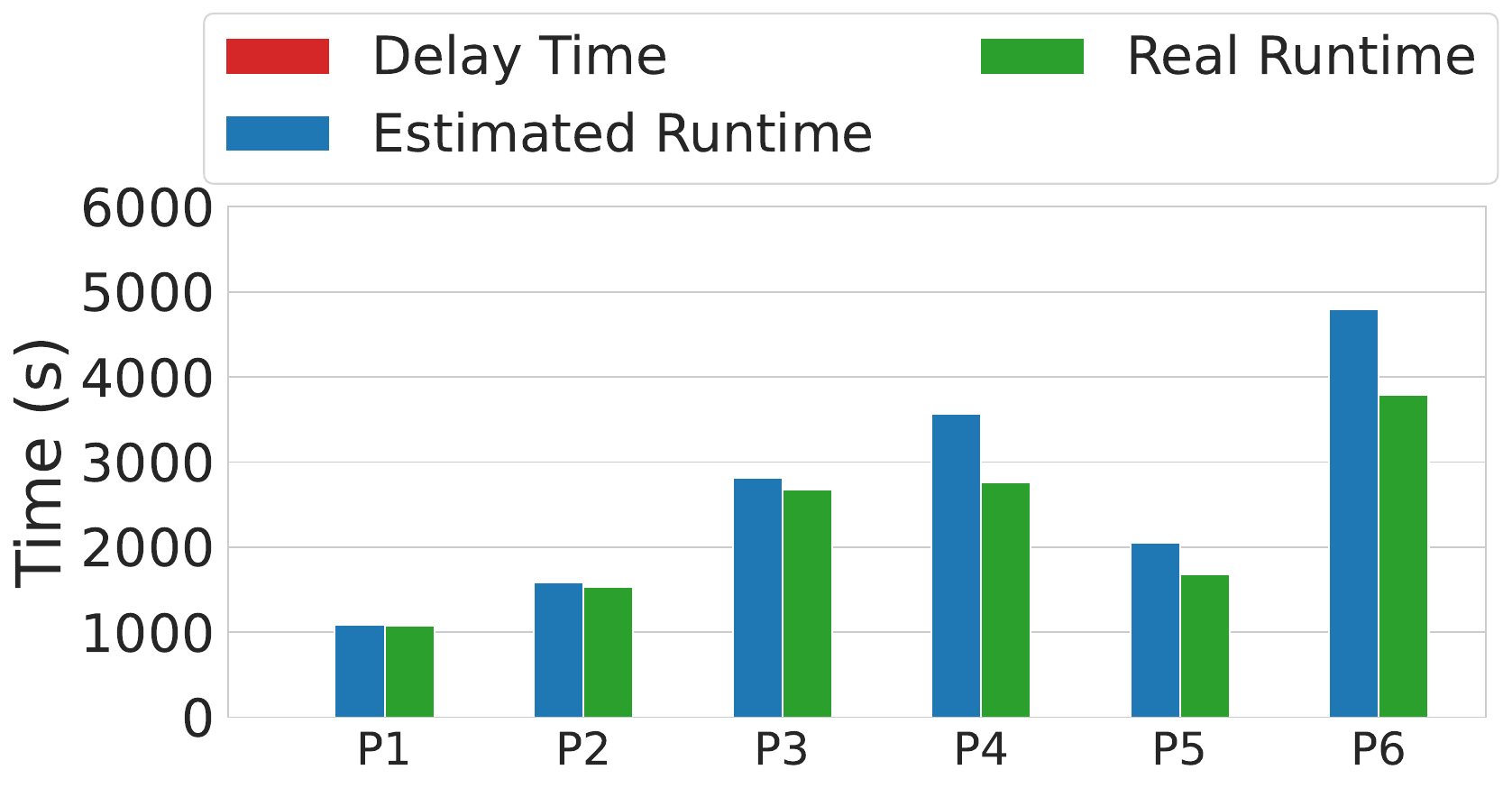}
    \label{fig:minimum_finish_time_performance}}
    \hspace{-4mm}
    \subfigure[Minimum Run Time (MRT)]{\includegraphics[keepaspectratio=true,angle=0,width=.32\textwidth]{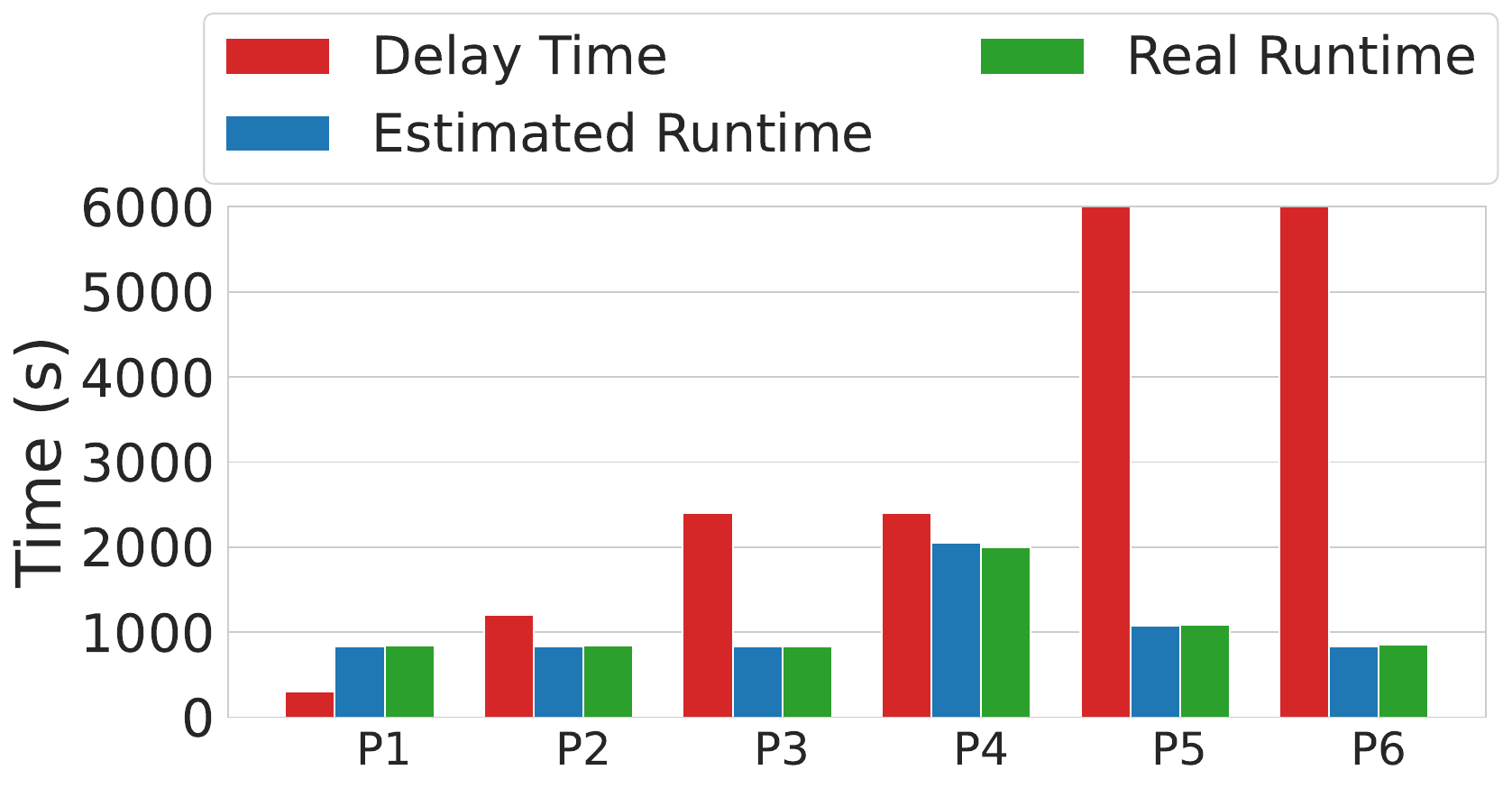}
    \label{fig:minimum_execution_tim_performance}}
    \hspace{-4mm}
    \subfigure[Weighted Average of MDT and MRT]
    {\includegraphics[keepaspectratio=true,angle=0,width=.32\textwidth]{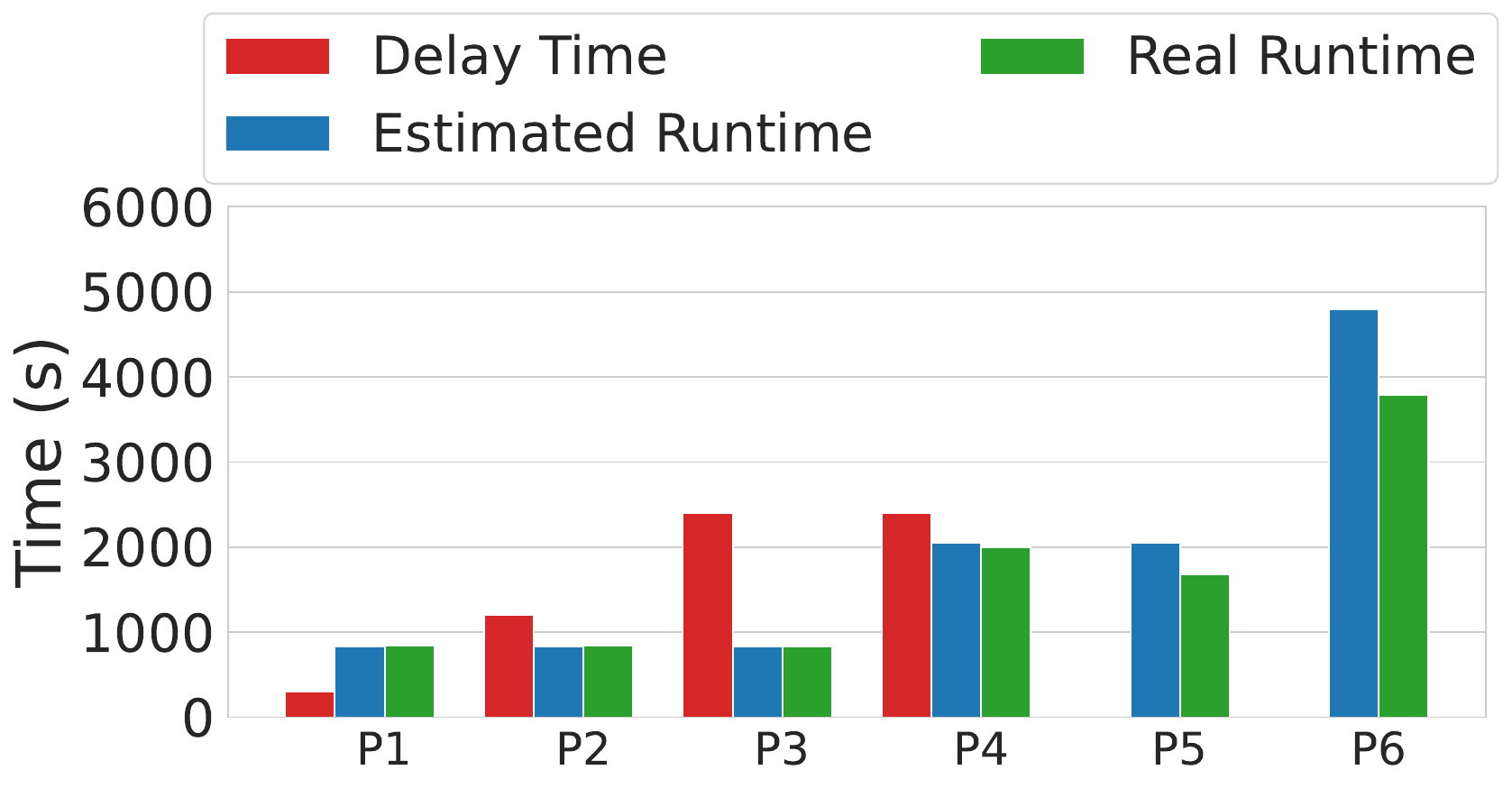}
    \label{fig:weighted_performance}}
    \hspace{-4mm}
\caption{Estimated and actual execution times for DLIO benchmark for various burst patterns. The estimated run times are mostly within $10\%$ of actual times.}
\vspace{-5mm}
\label{fig:prediction_and_real_comparision}
\end{center}
\end{figure*}

Next, we introduce a scheduling algorithm (Algorithm~\ref{alg:burst-awarea-sched})  to determine the start time for a job based on its sensitivity to bursts and the predicted burst pattern. The burst pattern returns whether or not an I/O burst is expected in the next two hours in five minutes intervals. Thus, it is a vector of size $24$, consisting of 0s (indicating no burst prediction) and 1s (indicating burst prediction). As an example, the burst pattern [0,0,0,1,1] means that no I/O burst is expected for the first three intervals (i.e., 15 minutes) while an I/O burst is expected in the subsequent two intervals (i.e., 10 minutes). The algorithm then calculates the estimated run time for different start times. Since the bursts are predicted in 5-minute intervals, the start times are defined at five minutes granularity. Moreover, since the prediction model can estimate up to two hours ahead, we evaluate $24$ different start time options ($24*5=120$ minutes = $2$ hours). If an application is expected to run beyond the prediction interval (i.e., two hours from the current time) either due to long runtime or late start time, we assume the worst and fill the rest of the intervals with burst prediction. It is presumed that applications' execution time with and without bursts can be calculated from the historical data.

\begin{figure*}[]
\begin{center}
    \subfigure[DLIO Benchmark]
    {\includegraphics[keepaspectratio=true,angle=0,width=.32\textwidth]{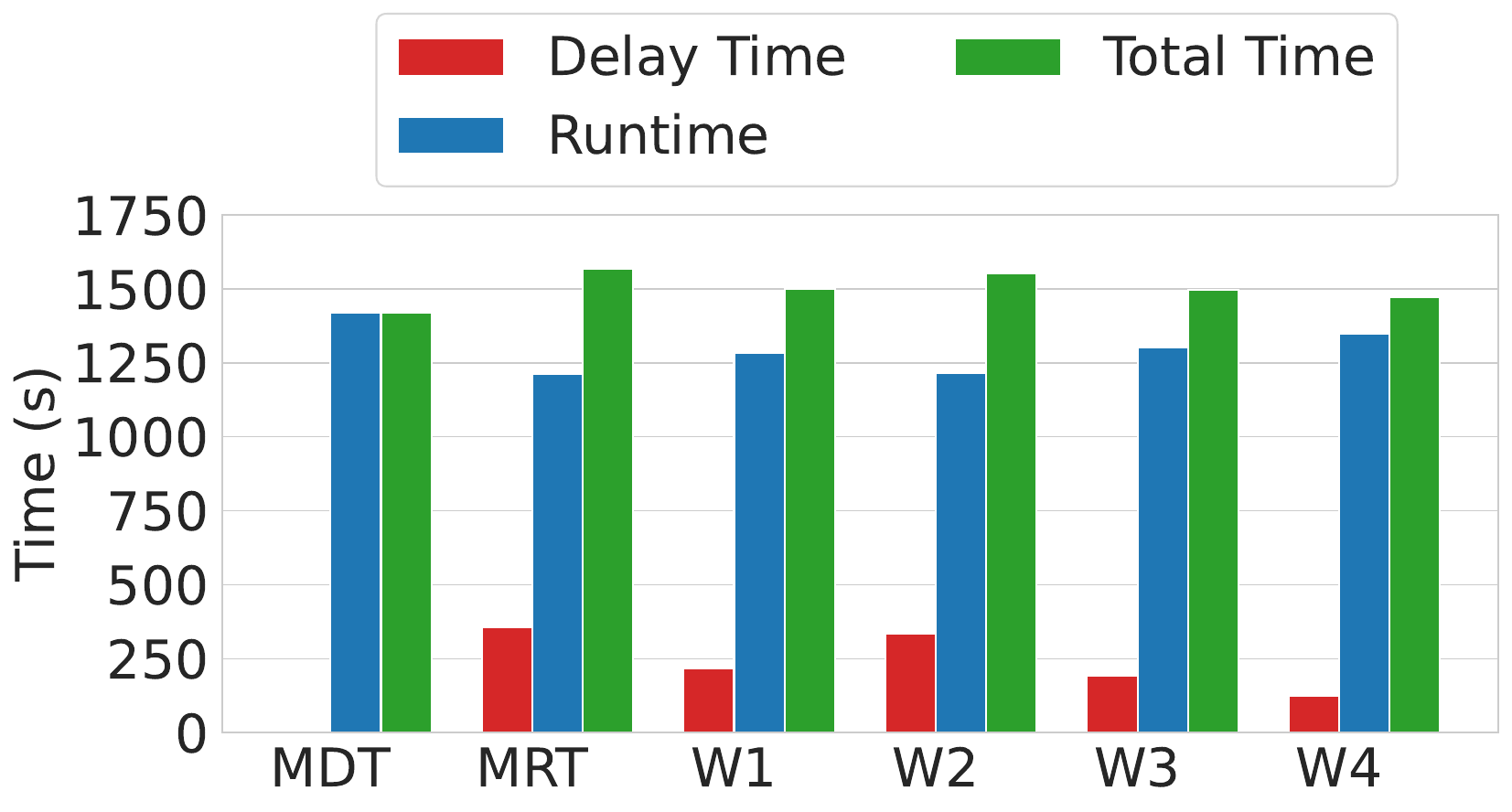}
    \label{fig:app1}}
    \hspace{-4mm}
    \subfigure[Application 2]{\includegraphics[keepaspectratio=true,angle=0,width=.32\textwidth]{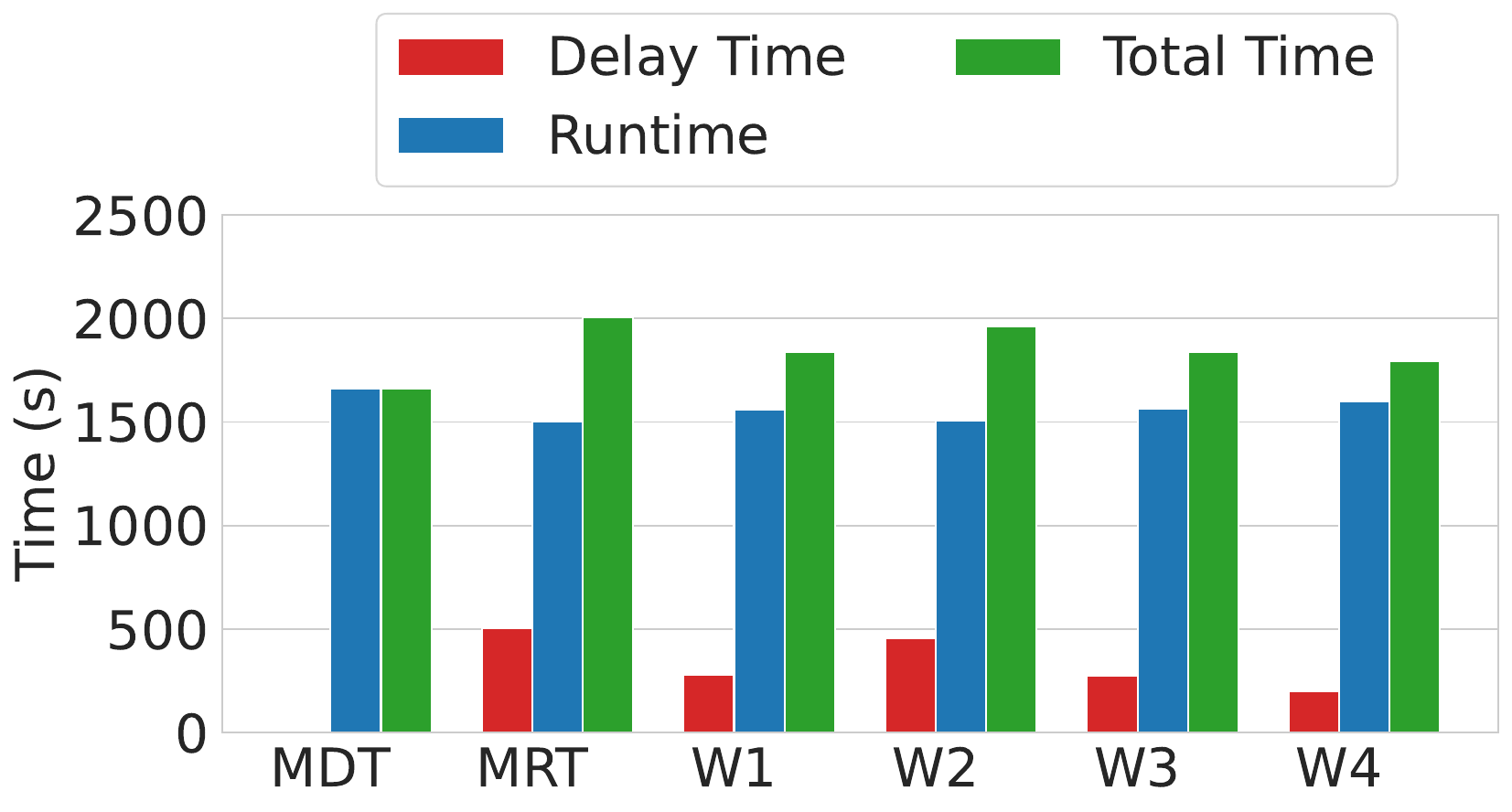}
    \label{fig:app2}}
    \hspace{-4mm}
    \subfigure[Application 3]{\includegraphics[keepaspectratio=true,angle=0,width=.32\textwidth]{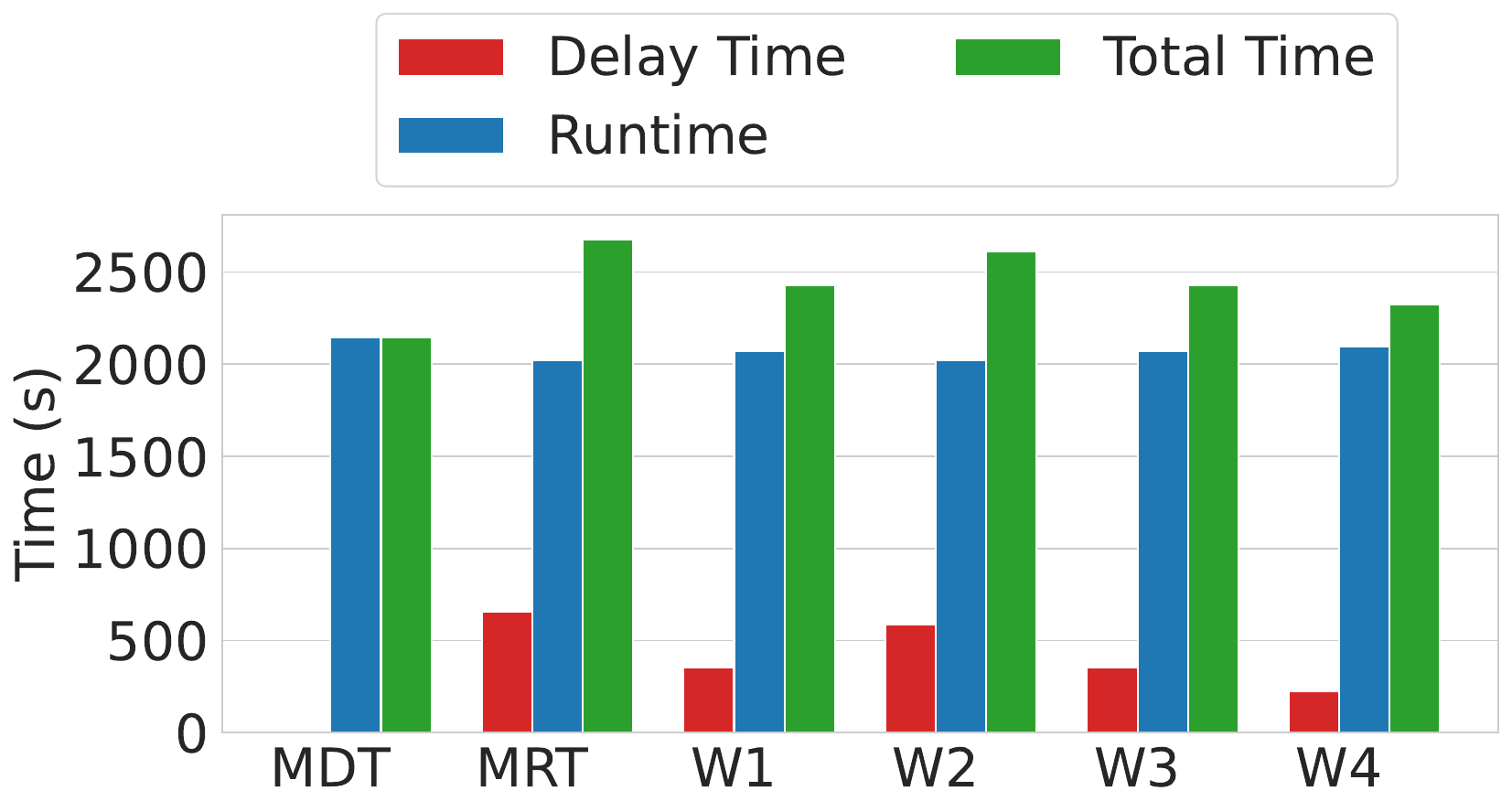}
    \label{fig:app3}}
    \hspace{-4mm}
\caption{Delay and run-time comparison for three different applications using all burst patterns in the Blue Waters dataset.}
\vspace{-8mm}
\label{fig:delay_run_total_time_prediction}
\end{center}
\end{figure*}

\begin{figure*}[]
\begin{center}
    \subfigure[DLIO Benchmark - Delay Time]
    {\includegraphics[keepaspectratio=true,angle=0,width=.32\textwidth]{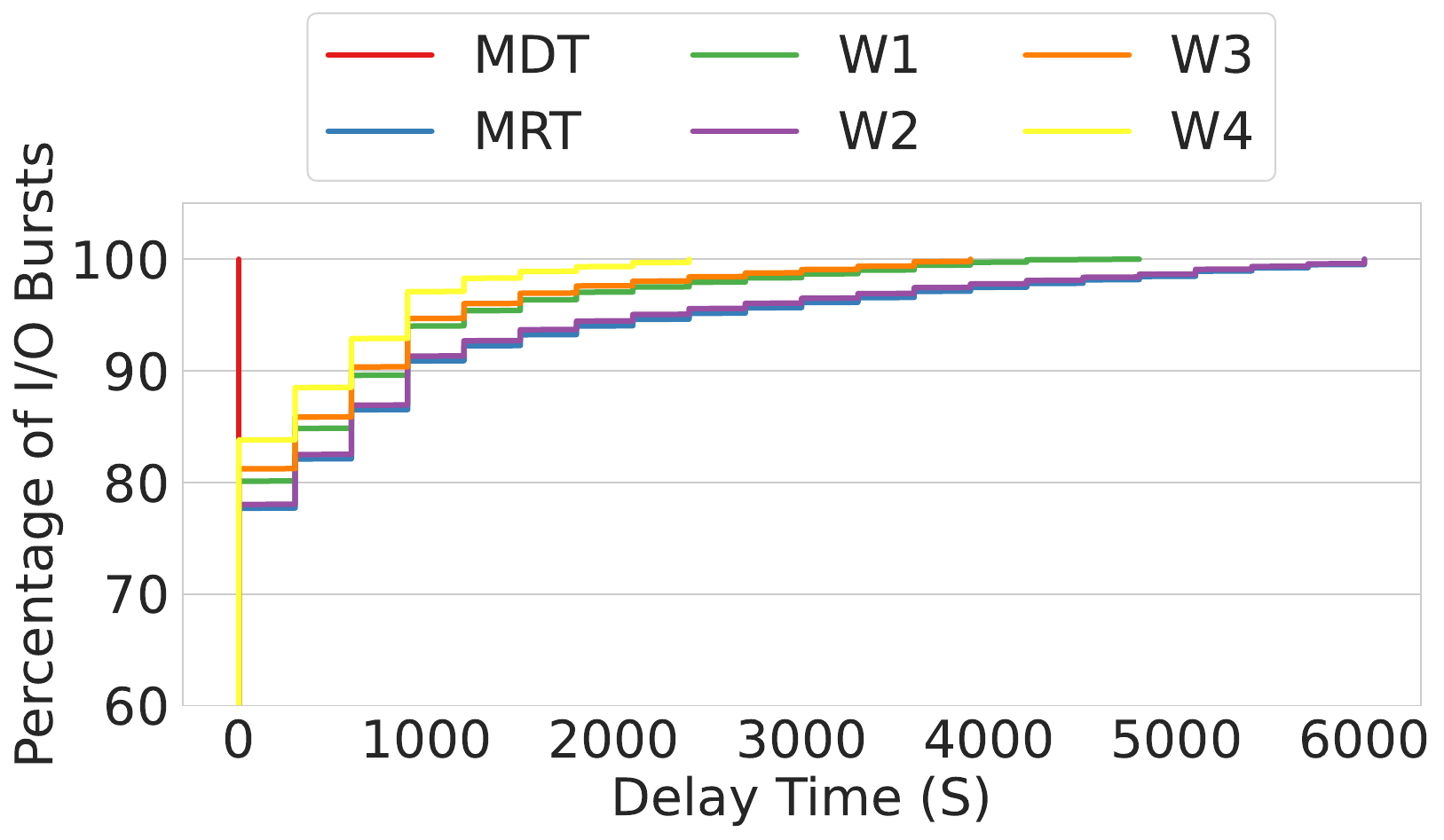}
    \label{fig:app1_delay}}
    \hspace{-4mm}
    \subfigure[DLIO Benchmark - Run Time]{\includegraphics[keepaspectratio=true,angle=0,width=.32\textwidth]{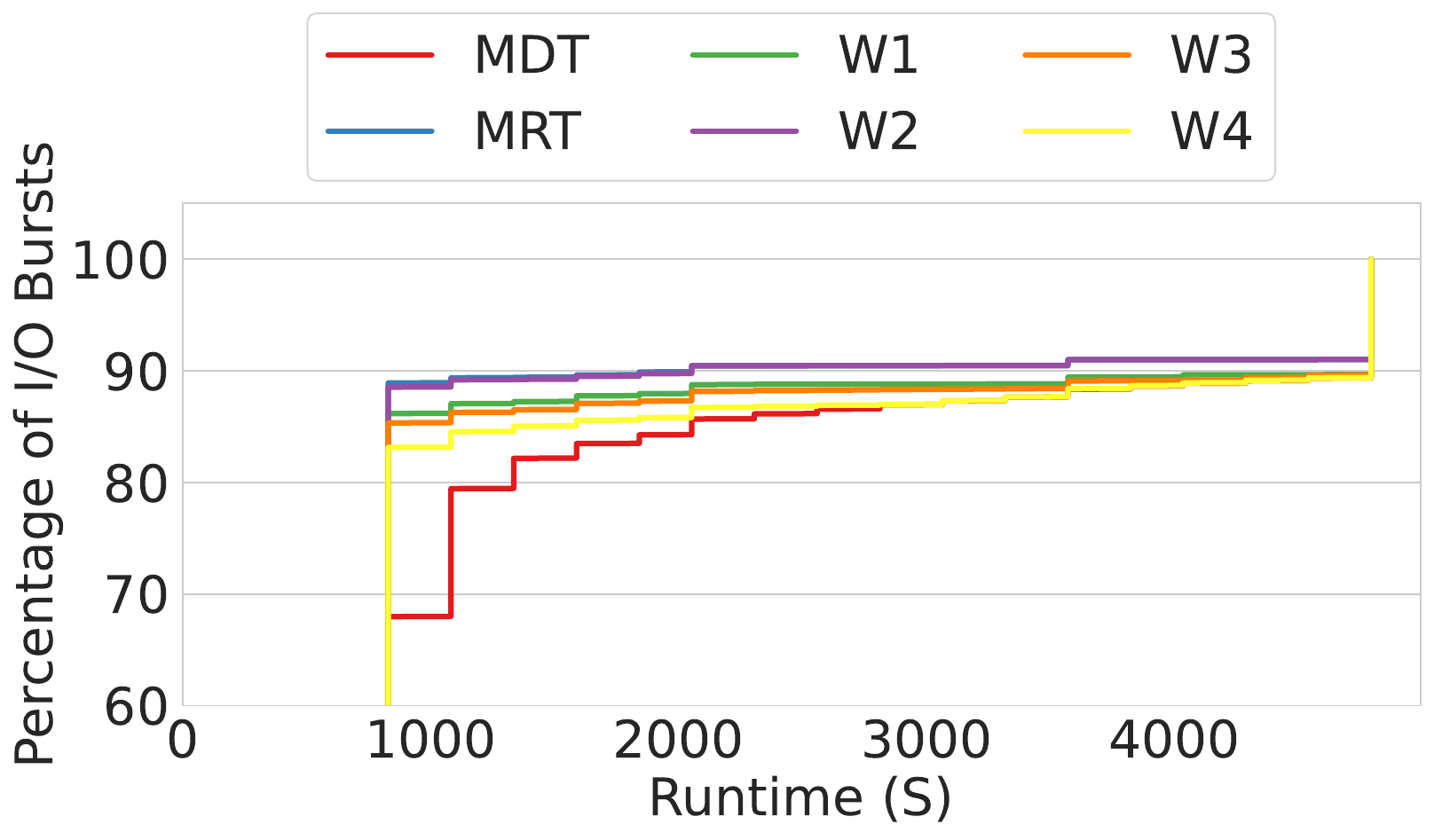}
    \label{fig:app1_run}}
    \hspace{-4mm}
    \subfigure[DLIO Benchmark - Total Time]{\includegraphics[keepaspectratio=true,angle=0,width=.32\textwidth]{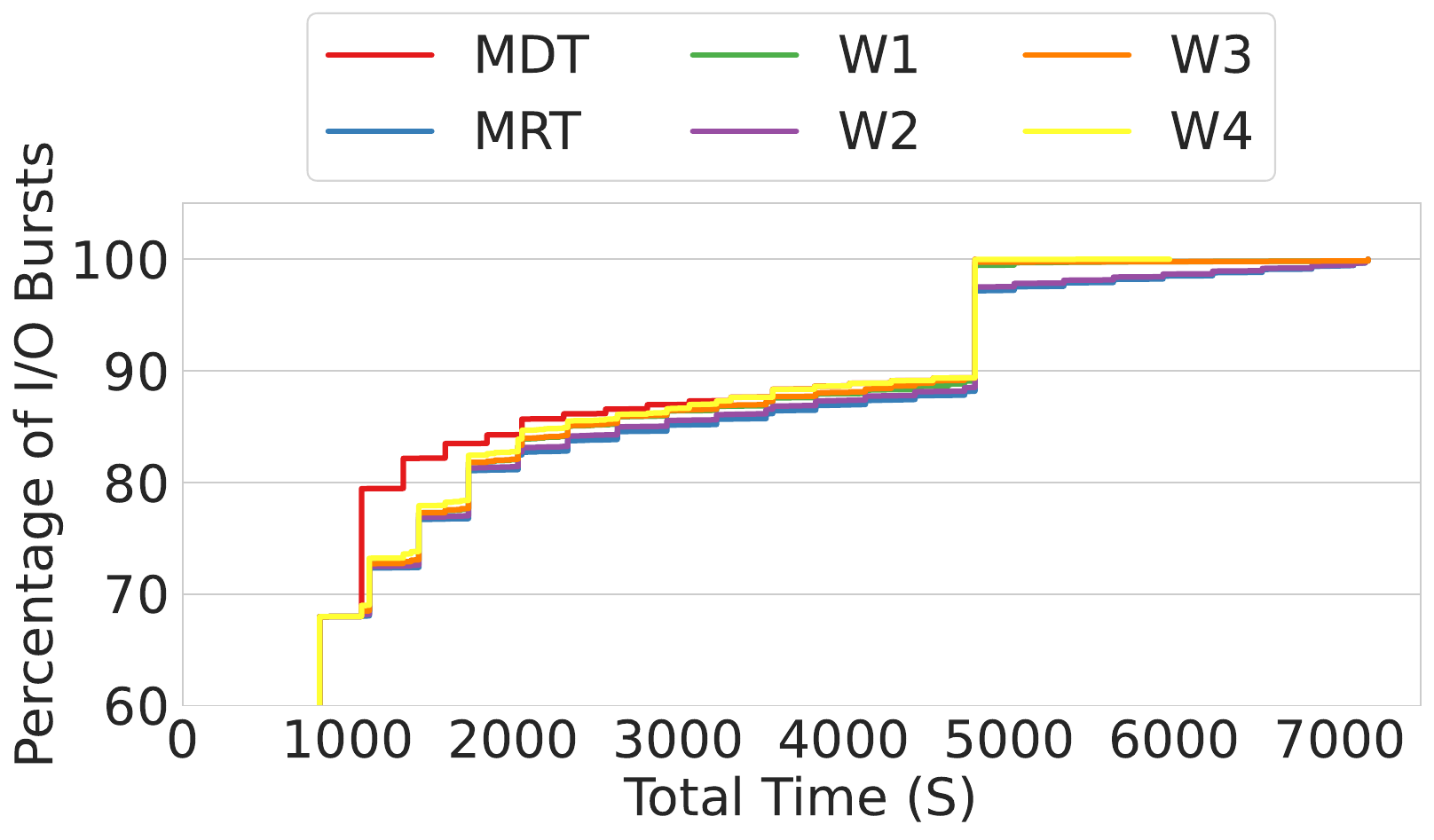}
    \label{fig:app1_total}}
    \hspace{-4mm}


\caption{CDF for the delay, runtime, and total time of DLIO benchmark using burst-aware scheduler for the Blue Waters dataset.}
\vspace{-8mm}
\label{fig:CDF_delay_run_total_time_prediction}
\end{center}
\end{figure*}

The algorithm takes minimum (without burst) and maximum (with burst) execution times for the application as an argument and calculates the amount of progress it can make in burst and non-burst intervals in lines ~\ref{line:progress_nb} and \ref{line:progress_b}. Then, it iterates over every possible start time option in line~\ref{line:iterate}. Since the prediction model can estimate burst patterns for $24$ intervals each five minutes long, we restrict the possible start time options to $24$ (i.e., $N=24$). In lines~\ref{line:time_estimate_start} to \ref{line:time_estimate_end}, we estimate the execution time for each possible start interval based on the burst pattern. Finally, we calculate a score using delay and run times. The weights used in the score calculation can be adjusted to meet specific objectives. On one end, jobs can be executed immediately regardless of burst predictions to minimize the delay and completion times (i.e., $\alpha=1$), which we refer to as \textit{Minimum Delay Time (MDT)}. On the other hand, jobs can be delayed to avoid the burst as much as possible such that the execution time can be minimized (i.e., $\alpha=0$), referred to as \textit{Minimum Run Time (MRT)}. The value of $\alpha$ can be tuned to strike a balance between MDT and MRT approaches.

Figure~\ref{fig:prediction_and_real_comparision} presents the estimated and actual run times of the DLIO benchmark under six different burst patterns sampled from the Blue Waters dataset. Actual run time is captured by running the DLIO benchmark and injecting background load as observed in each sampled burst pattern. We draw two conclusions based on these results. First, MRT can lower the execution time by as much as five times compared to MDT by delaying the start time to avoid I/O bursts. Second, the estimated run time of the application (as calculated by Algorithm~\ref{alg:burst-awarea-sched}) is mostly within $10\%$ of the actual runtime for all solutions and burst patterns though it deviates as much as $20\%$ in very few cases. We believe this is a sufficient approximation of actual runtime and run a large-scale simulation using all burst patterns (with at least one burst interval) observed in the Blue Waters dataset. Although $2^{24}$ different possible burst pattern combinations exist, we observed $2,751$ unique burst patterns and $17,269$ total intervals (consisting of $24$ 5-minute intervals) with as least one burst interval. We then run the Algorithm~\ref{alg:burst-awarea-sched} for all $17,269$ intervals.

Figure~\ref{fig:app1} shows the average delay, runtime, and total time for the DLIO benchmark when simulated under all burst patterns. We observe that while the application run time is around $1,400$ seconds when MDT is used, it decreases to around $1,200$ seconds with MRT. W1, W2, W3, and W4 represent $\alpha$ values of $0.7, 0.6, 0.5,$ and $0.4$. It is clear that one can reduce application execution time considerably in exchange for postponing the start time to avoid I/O bursts. Since different applications can have different sensitivities to I/O bursts, we also simulated two other applications whose minimum and maximum times are different from the DLIO benchmark. Specifically, the application in Figure~\ref{fig:app2} has a minimum execution time of $1,260$ and a maximum execution time of $3,780$. These values are $1,830$ and $3,660$ seconds in figure~\ref{fig:app3}, respectively. Despite the lower difference between minimum and maximum execution times (which refer to less sensitivity to I/O bursts), we still observe up to $30\%$ (with an average of $9\%$) decrease in runtime by delaying the start time by $5-30$ minutes. As HPC applications are charged based on their runtime, lowering runtime by postponing jobs slightly can result in significant cost savings for HPC users. Furthermore, one can tune the value of $\alpha$ to find a tradeoff between total time and execution time such that one can avoid delaying jobs too long for a marginal reduction in execution time. Figure ~\ref{fig:app1_delay}, ~\ref{fig:app1_run}, and~\ref{fig:app1_total} show the cumulative distribution of delay, runtime, and total time for the DLIO benchmark when simulated under all burst patterns. We observe that while all strategies are competitive in total time, MRT results in the maximum, and MDT always gets the minimum delay time. While it is possible to extend the Algorithm~\ref{alg:burst-awarea-sched} to consider the severity of bursts, we leave it as a future work.

\subsection{Limitations} 
Despite yielding high-precision prediction results, this work has a few limitations. First, the Darshan toolkit is unable to capture all I/O operations; thus, the overall system I/O is probably higher than what is captured by the Darshan logs. However, Darshan logs are commonly used to characterize I/O behavior in leadership supercomputers ~\cite{luu2015multiplatform,snyder2016modular,kim2020towards,luttgau2018toward,lockwood2019understanding}; hence we believe modeling system-level Darshan load is still helpful to estimate overall I/O pattern of most HPC applications. Second, we selected I/O burst thresholds for read and write operations separately to mark $1\%$ of all time intervals as ``burst''. However, it is possible that some of the time intervals in the burst class may not have sufficient I/O rate to be classified as system-level high I/O contention. Even if this is the case, we argue that the proposed models are still helpful in detecting a significant increase in system-level I/O which can be sufficient to degrade the performance of large-scale I/O intensive applications as I/O contention is expected to be higher in ``burst'' intervals compared to other intervals.

\section{Conclusion and Future Directions}
In this study, we processed Darshan logs from the three supercomputers to gain insights into system-level I/O rates in HPC clusters. We observed that both read and write operations exhibit significant fluctuations over time, causing bursts. We evaluated the performance of various machine learning models using multiple features in predicting I/O bursts ahead of time. Experimental results show that XGBoost models attain the best overall performance with more than $90\%$ F-1 scores both for read and write I/O burst predictions. We also find that it can attain more than $70\%$ accuracy when attempting to identify the severity level of the bursts. Finally, we demonstrate a potential use case of I/O burst prediction on the application runtime through a burst-aware job scheduling algorithm. The proposed algorithm can postpone the execution time of jobs if the burst condition is likely to improve in the near future, allowing applications to run as much as $5\times$ faster. In future work, we aim to propose adaptive threshold selection to automatically find ideal burst thresholds since the thresholds should be tailored to the system instead of using our arbitrary values. 




\section*{Acknowledgement}
The work in this study was supported in part by the NSF grants 2145742 and 2007789.

\bibliographystyle{IEEEtran}
\bibliography{main}

\end{document}